\documentclass{jpsj2}
\usepackage{txfonts}
\usepackage{bm}

\title{First-Principles Theory of Momentum Dependent Local Ansatz Approach to Correlated Electron System}

\author{Sumal Chandra\thanks{E-mail address: k138609@eve.u-ryukyu.ac.jp, be published in J. Phys. Soc. Jpn. $\bm{85}$ (2016).} and Yoshiro Kakehashi\thanks{yok@sci.u-ryukyu.ac.jp}}
\inst{Department of Physics and Earth Sciences, Faculty of Science,\\
University of the Ryukyus, \\
1 Senbaru, Nishihara, Okinawa, 903-0213, Japan} 

\abst{We have extended the momentum-dependent local-ansatz (MLA) wavefunction method to the first-principles version using the tight-binding LDA+U Hamiltonian for the description of correlated electrons in the real system. The MLA reduces to the Rayleigh-Schr\"odinger perturbation theory in the weak correlation limit, and describes quantitatively the ground state and related low-energy excitations in solids. The theory has been applied to the paramagnetic Fe. The role of electron correlations on the energy, charge fluctuations, amplitude of local moment, momentum distribution functions, as well as the mass enhancement factor in Fe has been examined as a function of Coulomb interaction strength. It is shown that the inter-orbital charge-charge correlations between $d$ electrons make a significant contribution to the correlation energy and charge fluctuations, while the intra-orbital and inter-orbital spin-spin correlations make a dominant contribution to the amplitude of local moment and the mass enhancement in Fe. Calculated partial mass enhancements are found to be 1.01, 1.01, and 3.33 for $s$, $p$, and $d$ electrons, respectively. The averaged mass enhancement 1.65 is shown to be consistent with the experimental data as well as the recent results of theoretical calculations.}

\kword{ first-principles theory, variational method, momentum-dependent local ansatz, Gutzwiller wevefunction, electron correlations, momentum distribution function, iron, transition metal }

\begin{document}
\maketitle

\section{Introduction}

The density functional theory (DFT) has been well developed in the past half century towards quantitative description of the properties of solids, and acts nowadays as a powerful tool for explaining the ground-state properties of materials and their electronic structure. In fact, the DFT based on the local density approximation (LDA) or generalized gradient approximation (GGA) explains many aspects in solids such as the cohesive properties, the Fermi surface in metals, and optical properties of metallic systems.\cite{jcs72, rmm12}

The DFT, however, is not sufficient to describe quantitatively the properties of more correlated electron systems. For example, it fails in explaining the reduction of the cohesive energy in 3d transition metals\cite{jfcm77}, the formation of a satellite peak in the X-ray photoemission spectroscopy (XPS) data of Ni\cite{drp79, ali81}, and the angle resolved photoemission spectroscopy (ARPES) data in Fe pnictides and cuprates. These properties cannot be understood without taking into account directly the many-body effects, $i.e.,$ electron  correlation effects.

In order to describe the many-body phenomena which cannot be explained by the band theory, various theories have been developed so far. These theories are based on the variational method, the Green function techniques, as well as the numerical techniques such as the exact diagonalization method and the quantum Monte-Carlo (QMC) technique. The dynamical mean field theory (DMFT) combined with the LDA+U Hamiltonian,\cite{gko06, via10} which is equivalent to the first-principles dynamical coherent potential approximation (DCPA) we developed\cite{ykap11, yka12}, is such an approach based on the Green function technique and the effective medium method. In this approach, we can replace the surrounding interactions with an effective medium and solve the impurity problem using various methods. The theory has been applied to many systems with strong electron correlations.  

The variational approach is the simplest and oldest methods to treat electron correlations at the ground state.\cite{pfu95, pfu12,pfu02} The trial wavefunction is chosen to include the minimum basis set with variational parameters. The Gutzwiller wavefunction (GW) \cite{mcg63, mcg64, mcg65} is one of the popular ansatz, and has been applied to a number of correlated electron systems such as Ni\cite{jbu03} and Fe pnictides \cite{gt10, ts12}. The method has been extended to the first-principles version on the basis of the LDA+U Hamiltonian.\cite{jbu11, jbu12} The first-principles GW theory has been applied to many systems, and clarified the physics of electron correlations such as the correlation effects on the magnetism, the heavyfermion behavior, and the metal-insulator transition. 

The Gutzwiller wavefunction however does not reduce to the second-order perturbation theory in the weak Coulomb interaction limit. Therefore it does not describe quantitatively the properties of correlated electron system. This is serious for the quantitative description of effective mass enhancement factor associated with the low energy excitations in the vicinity of the Fermi surface, because it is obtained by a renormalization of the counterpart in the weak Coulomb interaction limit according to the Fermi liquid theory. 
 
In order to overcome the difficulty, we recently proposed the momentum-dependent local ansatz (MLA) wavefunction which goes beyond the GW.\cite{ykt08, mar11, mar13} The MLA is an extension of the local ansatz approach (LA) in which the residual Coulomb interaction operators are used to expand the Hilbert space for describing electron correlations.\cite{gs77, gs78, gs80} In the MLA, we expand the Hilbert space by means of the two-particle excited states with momentum-dependent variational parameters in the momentum space, and project these states onto the local orbitals again. In this way, we can obtain more flexible correlated electron states. The theory overcomes the Gutzwiller wavefunction method and describes quantitatively the physical quantities associated with the low energy excitations such as the mass enhancement factor. In the next papers\cite{marp13, yk14}, we generalized the MLA introducing a hybrid wavefunction (HB) as a starting wavefunction, whose potential flexibly changes from the Hartree-Fock type to the alloy-analogy type by varying a weighting factor from zero to one. The HB-MLA can describe the correlated electron system from the weak to the strong Coulomb interaction regime, including the metal-insulator transition in infinite dimensions.

In this paper, we extend the MLA to the first-principles version on the basis of the tight-binding LDA+U Hamiltonian towards the quantitative description of correlated electron system. The first-principles MLA holds the high momentum and total-energy resolutions in the numerical calculations of the ground-state properties such as the momentum distribution function (MDF) and the ground-state energy because all the physical quantities in the MLA are expressed analytically and they are calculated with use of the Laplace transformation which transforms the 6-fold energy integrals into the 2-fold time integrals. We also point out that the present theory is the first which quantitatively explains the mass enhancement factor $m^{*}/m$ of Fe at zero temperature because most of the LDA+DMFT calculations for $m^{*}/m$ are limited to the finite temperature case \cite{aak10} and the zero-temperature calculations of the LDA+DMFT with use of the three-body theory failed in the quantitative explanation of experimental data of bcc Fe \cite{jsb09}. Furthermore the method has an advantage that it allows us to calculate any static physical quantities because we know the wavefunction itself.

Using the first-principles MLA, we examine the effects of the intra-orbital correlations, the inter-orbital charge-charge correlations, and the inter-orbital spin-spin correlations (, $i.e.,$ the Hund-rule correlations), on various quantities of the paramagnetic bcc Fe. We discuss the role of electron correlations on the correlation energy, charge fluctuations, amplitude of local moment, as well as mass enhancement factor. We demonstrate that the intra-orbital correlations and the inter-orbital charge-charge correlations make a significant contribution to the correlation energy and the suppression of charge fluctuations, while both the intra-orbital correlations and the inter-orbital spin-spin correlations cause the enhancement of the amplitude of local moment as well as the mass enhancement in Fe. We also show that the momentum distribution function strongly depends on the wave vector $\bm{k}$ because of the $d$ electron correlations. The calculated average mass enhancement factor $m^{*}/m=1.65$ is consistent with the experimental data obtained by the low-temperature specific heat\cite{wpe01, chc60, lch03} and the angle resolved photoemission spectroscopy (ARPES)\cite{jsb09}, as well as the recent result of the LDA+DMFT calculations\cite{aak10}. Preliminary results of the present work which were based on the lowest order calculations have been published as a proceedings\cite{sch15}.

In the following section, we present the first-principles MLA based on the tight-binding (TB) LDA+U Hamiltonian. We will introduce three kinds of correlators with the momentum dependent variational parameters, and construct the MLA wavefunction using them. We derive the correlation energy within the single-site approximation (SSA), and obtain the self-consistent equations for the momentum-dependent variational parameters. We also obtain the expressions for the charge fluctuations, the amplitude of magnetic moment, the momentum distribution function (MDF), as well as the mass enhancement factor. In \S 3, we present our numerical results of the calculations for the bcc Fe. We examine the ground-state correlation energy, the charge fluctuations, and the formation of amplitude of magnetic moment as a function of the Coulomb interaction strength. We also discuss the MDF, as well as the average mass enhancement of bcc Fe in comparison with those obtained by the other methods. We summarize our results in the last section, and discuss the remaining problems.

\section{Theory of the First-Principles MLA}
We consider the transition-metal system with an atom in the unit cell for simplicity, and adopt the first-principles LDA+U Hamiltonian, which is based on the tight-binding linear muffin-tin orbital method.\cite{via10, yka12} 
\begin{equation}
H=H_{1}+H_{2}.
\label{equ1}
\end{equation}
$H_{1}$ and $H_{2}$ denote the non-interacting and interacting parts of the Hamiltonian $H$. The former is given by
\begin{equation}
H_{1}=\sum_{iL\sigma}\epsilon^{0}_{L} \ \hat{n}_{iL\sigma}+\sum_{iLjL^{'}\sigma}{t}_{iLjL^{'}}\ a^{\dagger}_{iL\sigma}\,{a}_{jL^{'}\sigma}\, .
\label{equ2}
\end{equation}
Here $\epsilon^{0}_{L}$ is the atomic level of orbital $ L$ on site $i$. ${t}_{iLjL'}$  is the transfer integral between $iL$ and $jL'$. $L=(l, m)$ denotes the $s \,(l=0)$, $p\, (l=1)$, and $d \,(l=2)$ orbitals. $a^{\dagger}_{iL\sigma}{({a}_{iL\sigma})}$ is the creation (annihilation) operator for an electron on site $i$ with orbital $L$ and spin ${\sigma }$, and ${\hat{ n}_{iL\sigma}}=a^{\dagger}_{iL\sigma}{a}_{iL\sigma}$ is the number operator. The atomic level $\epsilon^{0}_{L}$ in $H_{1}$ is calculated from the LDA atomic level $\epsilon_{L}$ by subtracting the double counting potential as $\epsilon^{0}_{L} = \epsilon_{L} - \partial{E^{\mathrm{U}}_{\mathrm{ LDA}}}/\partial{n_{iL\sigma}}$. Here $n_{iL\sigma}$ is the charge density at the ground-state, $E^{\mathrm{U}}_{\mathrm{ LDA}}$ is a LDA functional for the  intra-atomic Coulomb interactions.\cite{via10} 
In the LDA+U Hamiltonian we assume that the $s p$ electrons are well described by the LDA in the band theory, and take into account only on-site Coulomb interactions between $d\, (l=2)$ electrons, so that the interaction part $H_{2}$ in Eq. (\ref{equ1}) is expressed as follows.
\begin{align}
H_{2}=\sum_{i}\Big[\sum_{m} {U}_{mm} \,\hat{n}_{ilm\uparrow}\,\hat{n}_{ilm\downarrow}+\sum_{(m,m')}\, \! \Big(U_{mm'}-\frac{1}{2}J_{mm'}\Big)\, \hat{n}_{ilm} \,\hat{n}_{ilm{'}}-2 \sum_{(m,m')}J_{mm'}\,\hat{\bm {s}}_{{ilm}}\cdot\hat{\bm {s}}_{ilm'}\Big]\,.
\label{equ3}
\end{align}
Here $U_{mm}\,(U_{mm'} )$ and $J_{mm'}$ denote the intra-orbital (inter-orbital) Coulomb and exchange interactions between $d$ electrons, respectively. $\hat{n}_{ilm}\, (\hat{\bm {s}}_{ilm})$ with $l = 2 $ is the charge (spin) density operator for $d $ electrons on site $i$ and orbital $m$. The operator  $\hat{\bm {s}}_{{iL}}$ is defined as $\hat{\bm {s}}_{{iL}}=\sum_{\gamma \gamma'} a^{\dagger}_{iL\gamma}(\bm{\sigma})_{\gamma\gamma'}\,{a}_{iL\gamma'}/2$. $\bm{\sigma}$ denotes the Pauli spin matrices.

In the first-principles MLA, we rewrite the Hamiltonian $H$ as the sum of the Hartree-Fock Hamiltonian $H_{0}$ and the residual interactions $H_{\mathrm {I}}$: 
\begin{align}
H=H_{0}+H_{\mathrm{I}}\,.
\label{equ4a}
\end{align}
The residual interaction part is given by
\begin{align}
H_{\mathrm{I}}&= \sum_{i}{\Big[\sum_{L}U_{LL}^{(0)}\ {O}^{(0)}_{iLL}+\sum_{(L,L')}U_{LL'}^{(1)} \ {O}^{(1)}_{iLL'}+\sum_{(L,L')}U_{LL'}^{(2)}\ {O}^{(2)}_{iLL'}\Big]}\,.  
\label{equ4}
\end{align}
The first term denotes the intra-orbital interactions, the second term is the inter-orbital charge-charge   interactions, and the third term expresses the inter-orbital spin-spin interactions, respectively. The Coulomb interaction energy parameters $U_{LL'}^{(\alpha)}$ are defined by $U_{LL}\delta_{LL'}$ $(\alpha=0)$, $U_{LL'}-J_{LL'}/2$ $(\alpha=1) $, and $-2J_{LL'}$ $ (\alpha=2)$, respectively. The two-particle operators ${O}^{(0)}_{iLL}$, ${O}^{(1)}_{iLL'}$, and ${O}^{(2)}_{iLL'}$  are defined by
\begin{equation}
{O}^{(0)}_{iLL}=\delta\hat{n}_{ilm\uparrow}\ \delta\hat{n}_{ilm\downarrow}\,,
\label{equ5}
\end{equation}
\begin{equation}
{O}^{(1)}_{iLL'}=\delta\hat{n}_{ilm}\ \delta\hat{n}_{ilm'}\,,
\label{equ6}
\end{equation}
\begin{equation}
{O}^{(2)}_{iLL'}=\delta\hat{\bm{s}}_{ilm}\cdot\delta\hat{\bm{s}}_{ilm'}\,.
\label{equ7}
\end{equation}
Note that $\delta A$ for an operator $A$ is defined by $\delta A=A-\langle A\rangle_{0}$, $\langle \sim \rangle_{0}$ being the average in the Hartree-Fock approximation.

As we have mentioned in the introduction, the LA makes use of the residual interactions $\lbrace{O}^{(\alpha)}_{iLL'}\rbrace$ as the correlators which expand the Hilbert space for correlated electrons. The LA however does not lead to the exact result in the weak Coulomb interaction limit. We introduce here the new correlators $\lbrace\tilde{O}^{(\alpha)}_{iLL'}\rbrace$ such that
\begin{align}
\tilde{O}^{(\alpha)}_{iLL'}&= \sum_{\{kn\sigma\}}\langle{k'_{2}n'_{2}\vert iL}\rangle_{\sigma'_{2}} \langle{iL\vert {k}_{2}{n}_{2}}\rangle_{\sigma_{2}} \langle{k'_{1}n'_{1}\vert iL'}\rangle_{\sigma'_{1}} \langle{iL'\vert {k}_{1}{n}_{1}}\rangle_{\sigma_{1}}  \nonumber \\
&\hspace{1cm}\times\lambda^{(\alpha)}_{{LL'}\{{2'2 1'1}\}}\ \delta(a^{\dagger}_{k'_{2}n'_{2}\sigma'_{2}}a_{{k}_{2}{n}_{2}\sigma_{2}})\ \delta(a^{\dagger}_{k'_{1}n'_{1}\sigma'_{1}}a_{{k}_{1}{n}_{1}\sigma_{1}})\,.
\label{equ8}
\end{align}
Here $\alpha$ denotes the three types operators $\alpha=$ 0, 1, and 2. $a^{\dagger}_{k n \sigma }{(a_{k n\sigma})}$ is the creation (annihilation) operator for an electron with momentum $\bm{k}$, band index $n$, and spin $\sigma $. They are given by those in the site representation as $a_{k n\sigma}=\sum_{iL}a_{iL\sigma}\langle k n\vert iL\rangle_{\sigma}$\,. 

The momentum dependent amplitudes $\lambda^{(\alpha)}_{{LL'}\{{2'2 1'1}\}}$ in Eq. (\ref{equ8}) are given by
\begin{equation}
\lambda^{(0)}_{{LL'}\{{2'2 1'1}\}}=\eta_{L k'_{2}n'_{2}k_{2}n_{2}k'_{1}n'_{1}k_{1}n_{1}}\ \delta_{LL'}\,\delta_{\sigma'_{2}\downarrow}\,\delta_{\sigma_{2}\downarrow}\,\delta_{\sigma'_{1}\uparrow}\,\delta_{\sigma_{1}\uparrow}\,,
\label{equ9}
\end{equation}
\begin{equation}
\hspace{-1.5cm}\lambda^{(1)}_{{LL'}\{2'2 1'1\}}=\zeta^{(\sigma_{2}\sigma_{1})}_{L L'k'_{2}n'_{2}k_{2}n_{2}k'_{1}n'_{1}k_{1}n_{1}}\ \delta_{\sigma'_{2}\sigma_{2}}\, \delta_{\sigma'_{1}\sigma_{1}}\,,
\label{equ10}
\end{equation}
\begin{align}
\lambda^{(2)}_{{LL'}\{{2'2 1'1}\}}&=\sum_{\sigma}\xi^{(\sigma)}_{L L'k'_{2}n'_{2}k_{2}n_{2}k'_{1}n'_{1}k_{1}n_{1}}\ \delta_{\sigma'_{2}-\sigma}\  \delta_{\sigma_{2}\sigma}\ \delta_{\sigma'_{1}\sigma}\ \delta_{\sigma_{1}-\sigma} \nonumber \\
&\hspace{1cm}+\frac{1}{2} \sigma_{1} \sigma_{2} \  \xi^{(\sigma_{2}\sigma_{1})}_{L L'k'_{2}n'_{2}k_{2}n_{2}k'_{1}n'_{1}k_{1}n_{1}}\ \delta_{\sigma'_{2}\sigma_{2}} \,\delta_{\sigma'_{1}\sigma_{1}}\,.                    
\label{equ11}
\end{align}
Here $\{2'21'1\}$ is defined by $\{2'21'1\}$=$k'_{2}n'_{2}\sigma'_{2}k_{2}n_{2}\sigma_{2}k'_{1}n'_{1}\sigma'_{1}k_{1}n_{1}\sigma_{1}$. $\eta_{L k'_{2}n'_{2}k_{2}n_{2}k'_{1}n'_{1}k_{1}n_{1}}$, $\zeta^{(\sigma_{2}\sigma_{1})}_{L L'k'_{2}n'_{2}k_{2}n_{2}k'_{1}n'_{1}k_{1}n_{1}}$, $\xi^{(\sigma)}_{L L'k'_{2}n'_{2}k_{2}n_{2}k'_{1}n'_{1}k_{1}n_{1}}$, and $\xi^{(\sigma_{2}\sigma_{1})}_{L L'k'_{2}n'_{2}k_{2}n_{2}k'_{1}n'_{1}k_{1}n_{1}}$ are variational parameters to be determined. It should be noted that $\tilde{O}^{(0)}_{iLL}$, $\tilde{O}^{(1)}_{iLL'}$, and $\tilde{O}^{(2)}_{iLL'}$ reduce to the local correlators, ${O}^{(0)}_{iLL}$, ${O}^{(1)}_{iLL'}$, and ${O}^{(2)}_{iLL'}$, respectively, when $\eta_{L k'_{2}n'_{2}k_{2}n_{2}k'_{1}n'_{1}k_{1}n_{1}}=\zeta^{(\sigma_{2}\sigma_{1})}_{L L'k'_{2}n'_{2}k_{2}n_{2}k'_{1}n'_{1}k_{1}n_{1}}=1$ and $\xi^{(\sigma)}_{L L'k'_{2}n'_{2}k_{2}n_{2}k'_{1}n'_{1}k_{1}n_{1}}=\xi^{(\sigma_{2}\sigma_{1})}_{L L'k'_{2}n'_{2}k_{2}n_{2}k'_{1}n'_{1}k_{1}n_{1}}=1/2$.

The two-particle correlators $\tilde{O}^{(0)}_{iLL}$, $\tilde{O}^{(1)}_{iLL'}$, and $\tilde{O}^{(2)}_{iLL'}$ describe the intra-orbital correlations, the inter-orbital charge-charge correlations, and the inter-orbital spin-spin correlations (, $i.e.,$ the Hund-rule correlations), respectively. Using the correlators $\lbrace\tilde{O}^{(\alpha)}_{iLL'}\rbrace$ and the Hartree-Fock ground-state wavefunction $\vert{\phi}\rangle$, we construct the first-principles MLA wavefunnction as follows.
\begin{equation}
\vert{\Psi}_\mathrm{MLA}\rangle={\Big[\prod_{i}{\Big( 1-\sum_{L}{\tilde{O}}^{(0)}_{iLL}-\sum_{(L,L')}{\tilde{O}}^{(1)}_{iLL'}-\sum_{(L,L')}{\tilde{O}}^{(2)}_{iLL'}\Big) } \Big]}\,\ \vert{\phi}\rangle\,.
\label{equ12}
\end{equation}

The variational parameters $\eta$'s, $\zeta$'s, and $\xi$'s in the correlators $\lbrace\tilde{O}^{(\alpha)}_{iLL'}\rbrace$ are obtained from the variational principle for the ground-state energy $E$.
\begin{align}
\langle H\rangle=\langle H\rangle_{0}+N\epsilon_c \geq E \,.
\label{equ13}
\end{align}
Here the correlation energy per atom $\epsilon_c$ is defined by $N\epsilon_c\equiv\langle \tilde{H} \rangle=\langle H \rangle-\langle H \rangle_{0}$. Note that $\tilde{H}\equiv H -\langle H \rangle_{0}=\tilde{H}_{0}+H_{I}$. $N$ is the number of atoms, and $\langle \sim \rangle$ denotes the full average with respect to $\vert \Psi_\mathrm{MLA}\rangle$.

We adopted the single-site approximation (SSA) to calculate the correlation energy $\epsilon_c$ (see Appendix A for the derivation). 
\begin{equation}
{\epsilon_c} =\frac{{- \langle {\tilde{O_i}^\dagger}} {H}_{I}\rangle_0 -\langle {H}_{I} \tilde{O_i}\rangle_0 
+\langle {\tilde{O_i}^\dagger} \tilde{H} \tilde{O_i}\rangle_0 }{1+\langle{\tilde{O_i}^\dagger\tilde{O_i}}\rangle_0}\,. 
\label{equ15}
\end{equation}
Here the operator $\tilde{O_i}$ is defined by $\tilde{O_i}=\sum_{L}{\tilde{O}}^{(0)}_{iLL}+\sum_{(L,L')}{\tilde{O}}^{(1)}_{iLL'}+\sum_{(L,L')}{\tilde{O}}^{(2)}_{iLL'}$. 

Each term in the correlation energy (\ref{equ15}) is calculated with use of Wick's theorem, and has the following form.
\begin{align}
\langle {H}_{I} \tilde{O_i}\rangle_0 
=\sum_{\alpha\alpha'}\sum_{<LL'>}\sum_{<L''L'''>}\sum_{\{kn\sigma\}}^{2'21'1}U_{LL'}^{(\alpha)}\ \lambda_{L''L'''\{2'21'1\}}^{(\alpha')}\ P_{LL'L''L'''}^{(\alpha\alpha')}(\{2'21'1\})\,,
\label{equ16}
\end{align}
\begin{equation}
\langle {\tilde{O_i}^\dagger} \tilde{H} \tilde{O_i}\rangle_0=\langle {\tilde{O_i}^\dagger} \tilde{H}_{0} \tilde{O_i}\rangle_0+\langle {\tilde{O_i}^\dagger} {H}_{I} \tilde{O_i}\rangle_0\,,
\label{equ17}
\end{equation}
\begin{align}
\langle {\tilde{O_i}^\dagger} \tilde{H}_{0} \tilde{O_i}\rangle_0
=\!\sum_{\alpha\alpha'}\sum_{<LL'>}\sum_{<L''L'''>}\sum_{\{kn\sigma\}}^{2'21'1}\sum_{\{k'n'\sigma'\}}^{4'43'3}\lambda_{LL'\{2'21'1\}}^{(\alpha)*}\,\lambda_{L''L'''\{4'43'3\}}^{(\alpha')}\,Q_{LL'L''L'''}^{(\alpha\alpha')}(\{2'21'1\}\{4'43'3\})\,,
\label{equ18}
\end{align}
\begin{align}
\langle {\tilde{O_i}^\dagger} {H}_{I} \tilde{O_i}\rangle_0
=\sum_{\alpha\alpha'}\sum_{<LL'>}\sum_{<L''L'''>}\sum_{\{kn\sigma\}}^{2'21'1}\sum_{\{k'n'\sigma'\}}^{4'43'3}\lambda_{LL'\{2'21'1\}}^{(\alpha)*}\,\lambda_{L''L'''\{4'43'3\}}^{(\alpha')}\,R_{LL'L''L'''}^{(\alpha\alpha')}(\{2'21'1\}\{4'43'3\})\,,
\label{equ19}
\end{align}
\begin{align}
\langle \tilde{O_i}^{\dagger} \tilde{O_i}\rangle_0 =\sum_{\alpha\alpha'}\sum_{<LL'>}\sum_{<L''L'''>}\sum_{\{kn\sigma\}}^{2'21'1}\sum_{\{k'n'\sigma'\}}^{4'43'3}\lambda_{LL'\{2'21'1\}}^{(\alpha)*} \lambda_{L''L'''\{4'43'3\}}^{(\alpha')} S_{LL'L''L'''}^{(\alpha\alpha')}(\{2'21'1\}\{4'43'3\})\,.
\label{equ20}
\end{align}
The sum $\sum_{<L L'>}$ in Eqs. (\ref{equ16})$\sim$(\ref{equ20}) is defined by $\sum_{L}$ when  $L'$=$L$, and by $\sum_{(L, L')}$ when $L'$$\neq$$L$. The explicit expressions of  $P_{LL'L'' L'''}^{(\alpha\alpha')}(\{2'21'1\}), Q_{ LL'L'' L'''}^{(\alpha\alpha')}(\{2'21'1\}\{4'43'3\})$, and $S_{LL'L'' L'''}^{(\alpha\alpha')}(\{2'21'1\}\{4'43'3\})$ are given in Appendix B (see Eqs. (\ref{equ2b}), (\ref{equ3b}), and (\ref{equ4b})). $R_{LL'L''L'''}^{(\alpha\alpha')}(\{2'21'1\}\{4'43'3\})$ are matrix elements related to the residual interactions and have the form $R_{LL'L''L'''}^{(\alpha\alpha')}(\{2'21'1\}\{4'43'3\})=\sum_{\alpha''}\sum_{<\bar{L}\bar{L'}>} U_{\bar{L}\bar{L'}}^{(\alpha'')}R_{LL'\bar{L}\bar{L'}L''L'''}^{(\alpha\alpha''\alpha')}(\{2'21'1\}\{4'43'3\})$.

We obtain the self-consistent equation from the stationary condition $\delta\epsilon_{c}=0$ as follows.
\begin{equation}
-\langle(\delta \tilde{O}_{i}^{\dagger}){H}_{I}\rangle_0+\langle (\delta\tilde{O}_{i}^{\dagger} )\tilde{H} \tilde{O_i}\rangle_0-\epsilon_c {\langle(\delta\tilde{O}_{i}^{\dagger}) \tilde{O_i}\rangle_0}+ c.c.= 0\,.
\label{equ21}
\end{equation}
The above condition yields the self-consistent equations for variational parameters. 
\begin{align}
&\sum_{\alpha'}\sum_{<L''L'''>}\sum_{\{kn\sigma\}}^{4'43'3}\Big[Q_{LL'L''L'''}^{(\alpha\alpha')}(\{2'21'1\}\{4'43'3\})\nonumber \\
&-\epsilon_{c}\, S_{LL'L''L'''}^{(\alpha\alpha')}(\{2'21'1\}\{4'43'3\})+\sum_{\alpha''}\sum_{<\bar{L}\bar{L'}>}U_{\bar{L}\bar{L'}}^{(\alpha'')}R_{LL'\bar{L}\bar{L'}L''L'''}^{(\alpha\alpha''\alpha')}(\{2'21'1\}\{4'43'3\})\Big]\ \lambda_{L''L'''\{4'43'3\}}^{(\alpha')}\nonumber \\
&=\sum_{\alpha'}\sum_{<L''L'''>}U_{L''L'''}^{(\alpha')}\ P_{L''L'''LL'}^{(\alpha'\alpha)*}(\{2'21'1\})\,.
\label{equ22}
\end{align}
The third term at the lhs (left-hand side) of Eq. (\ref{equ22}) is the higher order in the Coulomb interactions $U_{LL'}^{(\alpha)}$, so that we can neglect it and find the self-consistent solution in the weak Coulomb interaction limit as follows (see Appendix B for derivation).  
\begin{equation}
\lambda^{(\alpha)}_{{LL'}\{{2'2 1'1}\}}=\frac{C_{\sigma_{2}\sigma_{2}^{'}\sigma_{1}\sigma_{1}^{'}}^{(\alpha)}U_{LL'}^{(\alpha)}}{\Delta E_{k'_{2}n'_{2}\sigma'_{2} k_{2}n_{2}\sigma_{2} k'_{1}n'_{1}\sigma'_{1} k_{1}n_{1}\sigma_{1}}-\epsilon_c}\,.
\label{equ23}
\end{equation}
Here $\Delta E_{k'_{2}n'_{2}\sigma'_{2} k_{2}n_{2}\sigma_{2} k'_{1}n'_{1}\sigma'_{1} k_{1}n_{1}\sigma_{1}}$ is the two-particle excitation energy defined by $\Delta E_{k'_{2}n'_{2}\sigma'_{2} k_{2}n_{2}\sigma_{2} k'_{1}n'_{1}\sigma'_{1} k_{1}n_{1}\sigma_{1}}=\epsilon_{k'_{2}n'_{2}\sigma'_{2}}-\epsilon_{k_{2}{n}_{2}\sigma_{2}}+\epsilon_{k'_{1}n'_{1}\sigma'_{1}}-\epsilon_{k_{1}{n}_{1}\sigma_{1}}$. $C_{\sigma_{2}\sigma_{2}^{'}\sigma_{1}\sigma_{1}^{'}}^{(\alpha)}$  in the numerator is the coefficients given by
\begin{equation}
C_{\sigma_{2}\sigma_{2}^{'}\sigma_{1}\sigma_{1}^{'}}^{(\alpha)}=\begin{cases}
\delta_{\sigma'_{2}\downarrow}\,\delta_{\sigma_{2}\downarrow}\,\delta_{\sigma'_{1}\uparrow}\,\delta_{\sigma_{1}\uparrow} & (\alpha=0)\\
\delta_{\sigma'_{2}\sigma_{2}} \,\delta_{\sigma'_{1}\sigma_{1}} &  (\alpha=1)\\
\frac{1}{4} (\bm{\sigma})_{\sigma_{1}\sigma'_{1}}\cdot(\bm{\sigma})_{\sigma_{2}\sigma'_{2}}&  (\alpha=2)\,.
\end{cases}
\label{equ23ab}
\end{equation}

In order to obtain an approximate solution for more correlated electrons, we can assume the following solution, which interpolates between the weak Coulomb interaction limit and the atomic limit. 
\begin{equation}
\lambda^{(\alpha)}_{{LL'}\{{2'2 1'1}\}}=\frac{U_{LL'}^{(\alpha)}\sum_{\tau}C_{\tau\sigma_{2}\sigma_{2}^{'}\sigma_{1}\sigma_{1}^{'}}^{(\alpha)}\ \tilde{\lambda}_{\alpha\tau L L'}^{(\sigma_{2}\sigma_{1})}}{\Delta E_{k'_{2}n'_{2}\sigma'_{2} k_{2}n_{2}\sigma_{2} k'_{1}n'_{1}\sigma'_{1} k_{1}n_{1}\sigma_{1}}-\epsilon_c}\,.
\label{equ26}
\end{equation}
Here the spin-dependent coefficients $C_{\tau\sigma_{2}\sigma_{2}^{'}\sigma_{1}\sigma_{1}^{'}}^{(\alpha)}$ are defined by $C_{\sigma_{2}\sigma_{2}^{'}\sigma_{1}\sigma_{1}^{'}}^{(\alpha)}$  ($\alpha=0, 1$), $-(1/4)\ \ \sigma_{1}\sigma_{2}\delta_{\sigma'_{2}\sigma_{2}} \delta_{\sigma'_{1}\sigma_{1}}$ ($\alpha=2,\tau=l$), and $-(1/2)\sum_{\sigma}\delta_{\sigma'_{2}-\sigma}\delta_{\sigma_{2}\sigma}\delta_{\sigma'_{1}\sigma}\delta_{\sigma_{1}-\sigma}$ ($\alpha=2,\tau=t$), respectively. Note that $l\,(t)$ implies the longitudinal (transverse) component. The renormalization factors $\tilde{\lambda}_{\alpha\tau L L'}^{(\sigma\sigma')}$ are defined as $\tilde{\eta}_{LL'} \delta_{LL'}\delta_{\sigma'-\sigma}$ $(\alpha=0)$,  $\tilde{\zeta}_{LL'}^{(\sigma\sigma')}$ $(\alpha=1)$, $\tilde{\xi}_{tLL'}^{(\sigma)}\delta_{\sigma'-\sigma}$ $(\alpha=2, \tau=t)$, and $\tilde{\xi}_{lLL'}^{(\sigma\sigma')}$ $(\alpha=2,\tau=l)$, respectively. The renormalization factors $\tilde{\eta}_{LL}$, $\tilde\zeta^{(\sigma\sigma')}_{LL'}$, $\tilde \xi^{(\sigma)}_{tLL'}$, and $\tilde \xi^{(\sigma\sigma')}_{lLL'}$ are the new variational parameters to be determined.

Substituting Eq. (\ref{equ26}) into the elements in Eq. (\ref{equ15}), we obtain the following forms.
\begin{equation}
\langle {H}_{I} \tilde{O_i}\rangle_0=\sum_{\alpha\alpha'}\sum_{<LL'>}\sum_{<L''L'''>}U_{LL'}^{(\alpha)}\ U_{L''L'''}^{(\alpha')}\sum_{\tau\sigma\sigma'}\tilde{\lambda}_{\alpha'\tau L'' L'''}^{(\sigma\sigma')}\ P_{\tau LL'L'' L'''\sigma\sigma'}^{(\alpha\alpha')}\,,
\label{equ27}
\end{equation}
\begin{align}
\langle {\tilde{O_i}^\dagger} \tilde{H}_{0} \tilde{O_i}\rangle_0=\sum_{\alpha\alpha'}\sum_{<LL'>}\sum_{<L''L'''>}\!\!U_{LL'}^{(\alpha)}\ U_{L''L'''}^{(\alpha')}
\sum_{\tau\sigma\sigma'}\sum_{\tau'\sigma''\sigma'''}\!\tilde{\lambda}_{\alpha\tau L L'}^{(\sigma\sigma')*}\,\tilde{\lambda}_{\alpha'\tau' L'' L'''}^{(\sigma''\sigma''')}\,Q_{\tau\tau' LL'L'' L'''\sigma\sigma'\sigma''\sigma'''}^{(\alpha\alpha')}\,,
\label{equ28}
\end{align}
\begin{equation}
\langle {\tilde{O_i}^\dagger} {H}_{I} \tilde{O_i}\rangle_0=\sum_{\alpha}\sum_{<LL'>}U_{LL'}^{(\alpha)} \sum_{\tau\sigma\sigma'}\tilde{\lambda}_{\alpha\tau L L'}^{(\sigma\sigma')*}\  K_{\tau L L'\sigma\sigma'}^{(\alpha)}\,,
\label{equ29}
\end{equation}
\begin{align}
\langle {\tilde{O_i}^\dagger} \tilde{O_i}\rangle_0=\sum_{\alpha\alpha'}\sum_{<LL'>}\sum_{<L''L'''>}U_{LL'}^{(\alpha)}\ U_{L''L'''}^{(\alpha')}
\sum_{\tau\sigma\sigma'}\sum_{\tau'\sigma''\sigma'''}\tilde{\lambda}_{\alpha\tau L L'}^{(\sigma\sigma')*}\ \tilde{\lambda}_{\alpha'\tau' L'' L'''}^{(\sigma''\sigma''')}\ S_{\tau\tau' LL'L'' L'''\sigma\sigma'\sigma''\sigma'''}^{(\alpha\alpha')}\,.
\label{equ31}
\end{align}
Here the coefficients $P_{\tau LL'L'' L'''\sigma\sigma'}^{(\alpha\alpha')}, Q_{\tau \tau'LL'L'' L'''\sigma\sigma'\sigma''\sigma'''}^{(\alpha\alpha')}$, and $S_{\tau\tau'LL'L'' L'''\sigma\sigma'\sigma''\sigma'''}^{(\alpha\alpha')}$ are obtained by making use of Wick's theorem and the Laplace transformations. $K_{\tau L L'\sigma\sigma'}^{(\alpha)}$ in Eq. (\ref{equ29}) are the higher order corrections in the Coulomb interactions $\{U_{LL'}^{(\alpha)}\}$. They have the following form.
\begin{equation}
K_{\tau L L'\sigma\sigma'}^{(\alpha)}=\sum_{\alpha'}\sum_{<L''L'''>}\sum_{\tau'\sigma''\sigma'''}U_{L''L'''}^{(\alpha')}\ R_{\tau\tau' LL'L'' L'''\sigma\sigma'\sigma''\sigma'''}^{(\alpha\alpha')}\ \tilde{\lambda}_{\alpha'\tau' L'' L'''}^{(\sigma''\sigma''')}\,.
\label{equ30}
\end{equation}
The coefficients $R_{\tau\tau' LL'L'' L'''\sigma\sigma'\sigma''\sigma'''}^{(\alpha\alpha')}$ can be calculated again with use of Wick's theorem.

The self-consistent equations for the variational parameters $\tilde{\lambda}_{\alpha\tau' L L'}^{(\sigma\sigma')}$ are obtained from the stationary condition (\ref{equ21}) as follows. 
\begin{align}
&\sum_{\alpha'}\sum_{<L''L'''>}\sum_{\tau' \sigma''\sigma'''}U_{L''L'''}^{(\alpha')}\left(Q_{\tau\tau' LL'L'' L'''\sigma\sigma'\sigma''\sigma'''}^{(\alpha\alpha')}- \epsilon_c\ S_{\tau\tau' LL'L'' L'''\sigma\sigma'\sigma''\sigma'''}^{(\alpha\alpha')}\right)\ \tilde{\lambda}_{\alpha'\tau' L'' L'''}^{(\sigma''\sigma''')}\nonumber \\
&\hspace{5cm}=\sum_{\alpha'}\sum_{<L''L'''>} U_{L''L'''}^{(\alpha')}\ P_{\tau L'' L'''LL'\sigma\sigma'}^{(\alpha'\alpha)}-K_{\tau L L'\sigma\sigma'}^{(\alpha)}\,.
\label{equ32}
\end{align}
In the paramagnetic case, the variational parameters $\tilde{\lambda}_{\alpha\tau L L'}^{(\sigma\sigma')}$ are spin independent ($i.e., \tilde{\lambda}_{\alpha\tau L L'} $), and we can simplify Eq. (\ref{equ32}) as follows. 
\begin{equation}
\tilde{\lambda}_{0LL}=\tilde{Q}_{LL}^{-1}\left(P_{LL}-{U}_{LL}^{(0)-1}\ K_{LL}^{(0)}\right)\,,
\label{equ33}
\end{equation}
\begin{equation}
\tilde{\lambda}_{1LL'}=\tilde{Q}_{LL'}^{-1}\left(P_{LL'}-\frac{1}{4}\ {U}_{LL'}^{(1)-1}\ \bar{K}_{LL'}^{(1)}\right)\,,
\label{equ34}
\end{equation}
\begin{equation}
\tilde{\lambda}_{2lLL'}=-\tilde{Q}_{LL'}^{-1}\left(P_{LL'}+4\ {U}_{LL'}^{(2)-1}\ \bar{K}_{lLL'}^{(2)}\right)\,,
\label{equ35}
\end{equation}
\begin{equation}
\tilde{\lambda}_{2tLL'}=-\tilde{Q}_{LL'}^{-1}\left(P_{LL'}+4\ {U}_{LL'}^{(2)-1}\ K_{tLL'}^{(2)}\right)\,.
\label{equ36}
\end{equation}
Here $\tilde{Q}_{LL'}={Q}_{LL'}-{\epsilon_c} S_{LL'}$, $K_{LL'}^{(0)}=K_{LL'\downarrow\uparrow}^{(0)}$, $\bar{K}_{LL'}^{(1)}=\sum_{\sigma\sigma'}{K}_{LL'\sigma\sigma'}^{(1)}$, $\bar{K}_{lLL'}^{(2)}=\sum_{\sigma\sigma'}{K}_{lLL'\sigma\sigma'}^{(2)}$, and $K_{tLL'}^{(2)}=K_{tLL'\sigma-\sigma}^{(2)}$. The final expressions of all the elements are given in Appendix C.

Taking the same steps as in the derivation of $\epsilon_{c}$ in Eq. (\ref{equ15}) (see Appendix A), we can obtain the average of an operator $\tilde{A}= A-\langle A\rangle_{0}$ in the SSA as follows.
\begin{align}
\langle\tilde{A}\rangle=\sum_{i}\frac{{- \langle {\tilde{O_i}^\dagger}} \tilde{A} \rangle_0- \langle \tilde{A} \tilde{O_i}\rangle_0 
+\langle {\tilde{O_i}^\dagger} \tilde{A} \tilde{O_i}\rangle_0 }{1+\langle{\tilde{O_i}^\dagger\tilde{O_i}}\rangle_0}\,.
\label{equ14}
\end{align}
Using the formula (\ref{equ14}), we can obtain other physical quantities. The Fermi level $\epsilon_{F}$ is determined from the conduction electron number $n_{e}$ via the relation, 
\begin{equation}
n_{e}=\sum_{L}\langle n_{iL}\rangle\,.
\label{equ40a}
\end{equation}
Here and hereafter we omit the hat of the number operator for simplicity. The partial electron number of orbital $L$ on site $i$ is expressed as follows. 
\begin{equation}
\langle n_{iL}\rangle=\langle n_{iL}\rangle_{0}+\langle \tilde{n}_{iL}\rangle\,.
\label{equ40}
\end{equation}
Here $\langle n_{iL}\rangle_{0}$ denotes the  Hartree-Fock electron number. The correlation correction $\langle \tilde{n}_{iL}\rangle$ is obtained by using the formula (\ref{equ14}). 
\begin{align}
\langle \tilde{n}_{iL}\rangle=\frac{\langle\tilde{O}_{i}^{\dagger}\tilde{n}_{iL}\tilde{O}_{i}\rangle_{0}}{1+\langle{\tilde{O_i}^\dagger\tilde{O_i}}\rangle_0}\,.
\label{equ41}
\end{align}
The explicit expression of the numerator at the rhs (right-hand-side) of Eq. (\ref{equ41}) is given in Appendix D. 

It should be noted that the rhs of Eqs. (\ref{equ33}) $\sim$ (\ref{equ36}) contain the correlation energy $\epsilon_{c}$, the Fermi level $\epsilon_{F}$, as well as the variational parameters $\{\tilde{\lambda}_{\alpha\tau LL'}\}$. Thus Eqs. (\ref{equ15}), (\ref{equ33}) $\sim$ (\ref{equ36}), and (\ref{equ40}) have to be solved self-consistently.

The local charge fluctuation and the amplitude of the local moment for $d$ electrons are calculated from the following relations. 
\begin{align}
\langle (\delta n_{d})^{2}\rangle &=\sum_{L\sigma}^{d}\langle n_{iL\sigma}\rangle_{0}\ (1-\langle n_{iL\sigma}\rangle_{0})+\sum_{L\sigma}^{d}\langle \tilde{n}_{iL\sigma}\rangle\ (1-2\langle n_{iL\sigma}\rangle_{0}) \nonumber \\
&\hspace{.5cm}-\langle\tilde{n}_{id}\rangle^{2}+2\sum_{L}^{d}\langle O_{iLL}^{(0)}\rangle+2\sum_{(L,L')}^{d}\langle O_{iLL'}^{(1)}\rangle \,,
\label{equ42}
\end{align}
\begin{align}
\langle \bm{S}^{2}\rangle=&\frac{3}{4}\sum_{L\sigma}^{d}\langle n_{iL\sigma}\rangle_{0}\ (1-\langle n_{iL\sigma}\rangle_{0})+\frac{3}{4}\sum_{L\sigma}^{d}\langle \tilde{n}_{iL\sigma}\rangle\ (1-2\langle n_{iL-\sigma}\rangle_{0})\nonumber \\
&\hspace{.5cm}-\frac{3}{2}\sum_{L}^{d}\langle O_{iLL}^{(0)}\rangle+2\sum_{(L,L')}^{d}\langle O_{iLL'}^{(2)}\rangle \,.
\label{equ43}
\end{align}
Here the first terms at the rhs denote the Hartree-Fock contributions. $\langle\tilde{n}_{iL\sigma}\rangle$ in the second term is given by Eq. (\ref{equ41}) in which $\tilde{n}_{iL}$ has been replaced by $\tilde{n}_{iL\sigma}$, and is equal to $\langle\tilde{n}_{iL}\rangle/2$ in the paramagnetic state. $\langle\tilde{n}_{id}\rangle$ in the third term is defined by $\sum_{L}^{d}\langle\tilde{n}_{iL}\rangle$. The remaining correlation corrections at the rhs of Eqs. (\ref{equ42}) and (\ref{equ43}) are obtained from the residual interaction elements $\langle O_{iLL'}^{(\alpha)}\rangle$ using the formula (\ref{equ14}).  
\begin{equation}
\sum_{<LL'>}\langle O_{iLL'}^{(\alpha)}\rangle =\frac{\displaystyle{- \sum_{<LL'>}\langle {\tilde{O_i}^\dagger}} {O}_{iLL'}^{(\alpha)} \rangle_0- \sum_{<LL'>}\langle {O}_{iLL'}^{(\alpha)} \tilde{O_i}\rangle_0 
+\sum_{<LL'>}\langle {\tilde{O_i}^\dagger} {O}_{iLL'}^{(\alpha)} \tilde{O_i}\rangle_0 }{1+\langle{\tilde{O_i}^\dagger\tilde{O_i}}\rangle_0}\,. 
\label{equ45}
\end{equation}
The explicit expressions are summarized in Appendix E.

The momentum distribution function (MDF) is given as follows according to the formula (\ref{equ14}).
\begin{equation}
\langle n_{kn\sigma}\rangle=f(\tilde{\epsilon}_{kn\sigma})+\frac{N\langle\tilde{O}_{i}^{\dagger}\tilde{n}_{kn\sigma}\tilde{O}_{i}\rangle_{0}}{1+\langle{\tilde{O_i}^\dagger\tilde{O_i}}\rangle_0}\,.
\label{equ46}
\end{equation}
The first term at the rhs is the MDF for the Hartree-Fock independent electrons, which is given by the Fermi distribution function at zero temperature $f(\tilde{\epsilon}_{kn\sigma})$. $\tilde{\epsilon}_{kn\sigma}$ is the Hartree-Fock one-electron energy measured from the Fermi level $\epsilon_{F}$. The second term at the rhs of Eq. (\ref{equ46}) is the correlation corrections, where $\tilde{n}_{kn\sigma}$ is defined by $\tilde{n}_{kn\sigma}= n_{kn\sigma}-\langle n_{kn\sigma}\rangle_{0}$. The numerator is expressed as follows.
\begin{equation}
N\langle\tilde{O}_{i}^{\dagger}\tilde{n}_{kn\sigma}\tilde{O}_{i}\rangle_{0}=\sum_{\alpha\tau \ < LL'>} q_{\tau}^{(\alpha)}\ U_{LL'}^{(\alpha)2}\ \tilde{\lambda}_{\alpha\tau LL'}^{2}\ \Big(\hat{B}_{LL'n\sigma} (\bm{k})\ f(-\tilde{\epsilon}_{kn\sigma})-\hat{C}_{LL'n\sigma} (\bm{k})\ f(\tilde{\epsilon}_{kn\sigma})\Big)\,.
\label{equ46a}
\end{equation}
Here $q_{\tau}^{(\alpha)}$ is a constant factor taking the value 1 for $\alpha$=$0$, 2 for $\alpha$=$1$, 1/8 for $\alpha$=$2$, $\tau$=$l$, and 1/4 for $\alpha$=$2$, $\tau$=$t$, respectively. $\hat{B}_{LL'n\sigma} (\bm{k})$ is a momentum-dependent particle contribution above $\epsilon_{F}$ and is expressed as follows. 
\begin{equation}
\hat{B}_{LL'n\sigma} (\bm{k})=\vert u_{Ln\sigma}(\bm{k})\vert^{2}{B}_{L'L\sigma}({\epsilon}_{kn\sigma})+ \vert u_{L'n\sigma}(\bm{k})\vert^{2}{B}_{LL'\sigma}({\epsilon}_{kn\sigma})\,,
\label{equ46b}
\end{equation}
where $\{u_{Ln\sigma}(\bm{k})\}$ are the eigenvectors for a given $\bm{k}$ point. The hole contribution $\hat{C}_{LL'n\sigma} (\bm{k})$ is defined by Eq. (\ref{equ46b}) in which the energy dependent terms ${B}_{LL'\sigma} ({\epsilon}_{kn\sigma})$ have been replaced by ${C}_{LL'\sigma} ({\epsilon}_{kn\sigma})$. These are given by the Laplace transformation of the local density of states in the Hartree-Fock approximation. Their explicit expressions in the paramagnetic state are given in Appendix F (see Eqs. (\ref{eque4}), and (\ref{eque5})).

The quasiparticle weight $Z_{{k_{F}}n}$ characterizes the low energy excitations in metals. It is obtained by taking the difference between $\langle n_{kn\sigma}\rangle$ below and above the Fermi level $\epsilon_{F}$. Taking average over the Fermi surface, we obtain the average quasiparticle weight $Z$.
\begin{equation}
Z=1+\frac{\overline{\delta(N\langle\tilde{O}_{i}^{\dagger}\tilde{n}_{kn\sigma}\tilde{O}_{i}\rangle_{0})_{k_{F}}}}{1+\langle{\tilde{O_i}^\dagger\tilde{O_i}}\rangle_0}\,.
\label{equ46ab}
\end{equation}
Here the first term at the rhs denotes the Hartree-Fock part. The second term at the rhs is the correlation corrections. The upper bar in the numerator denotes the average over the Fermi surface, and ${\delta(N\langle\tilde{O}_{i}^{\dagger}\tilde{n}_{kn\sigma}\tilde{O}_{i}\rangle_{0})_{k_{F}}}$ means the amount of jump at the wavevector $\bm{k}_{F}$ on the Fermi surface. The explicit expression of ${\overline{\delta(N\langle\tilde{O}_{i}^{\dagger}\tilde{n}_{kn\sigma}\tilde{O}_{i}\rangle_{0})_{k_{F}}}}$ is given in Appendix F (see Eq. (\ref{eque11})).

In order to clarify the role of $s$, $p$, and $d$ electrons, we consider here the projected MDF for orbital $L$ defined by $\langle n_{kL\sigma}\rangle=\sum_{n}\langle n_{kn\sigma}\rangle\vert u_{Ln\sigma}(\bm{k})\vert^{2}$. Furthermore, we replace the energy $\epsilon_{kn\sigma}$ in the expression with $\epsilon_{kL\sigma}=\sum_{n}\epsilon_{kn\sigma}\vert u_{Ln\sigma}(\bm{k})\vert^{2}$, $i.e.,$ a common energy band projected onto the orbital $L$. We have then  
\begin{equation}
\langle n_{kL\sigma}\rangle=f(\tilde{\epsilon}_{kL\sigma})+\frac{N\langle\tilde{O}_{i}^{\dagger}\tilde{n}_{kL\sigma}\tilde{O}_{i}\rangle_{0}}{1+\langle{\tilde{O_i}^\dagger\tilde{O_i}}\rangle_0}\,.
\label{equ47}
\end{equation}
We can define the quasiparticle weight $Z_{L}$ for the electrons  with orbital symmetry $L$ by the jump of $\langle n_{kL\sigma}\rangle$ on the Fermi surface $\epsilon_{F}$. 
\begin{align}
Z_{L}=1+\frac{\overline{\delta(N\langle\tilde{O}_{i}^{\dagger}\tilde{n}_{kL\sigma}\tilde{O}_{i}\rangle_{0})}_{k_{F}}}{1+\langle{\tilde{O_i}^\dagger\tilde{O_i}}\rangle_0}\,.
\label{equ48}
\end{align}
It should be noted that the projected MDF depend on the momentum $\bm{k}$ only via $\tilde{\epsilon}_{kL\sigma}$. The explicit expressions of the correlation corrections at the rhs of Eqs. (\ref{equ47}) and (\ref{equ48}) are given in Appendix F (see Eqs. (\ref{eque17}), and (\ref{eque18})). Moreover we can verify the sum rule, 
\begin{align}
Z=\frac{1}{D}\sum_{L}Z_{L}\,.
\label{equ49}
\end{align}
Here $D$ is the number of orbitals ($D=9$ in the present case). The relation allows us to interpret $Z_{L}$ as a partial quasiparticle weight for the electrons with orbital $L$.

\section{Numerical Results for BCC Iron}

The bcc Fe has extensively been investigated theoretically with use of the realistic Hamiltonians with $s$, $p$, and $d$ orbitals at the ground states and at finite temperatures \cite{xy09, omi08, ail01, yka11, mni13}. But quantitative aspects on the physical properties of Fe have not yet been fully clarified even at the ground state. We performed numerical calculations for the paramagnetic Fe in order to clarify the quantitative aspects of the first-principles MLA and the effects of electron correlations in the properties of Fe.

The transfer integrals and the atomic level have been calculated with use of the Stuttgart tight-binding LMTO (linear muffin-tin orbital) package and the LDA+U scheme. We adopted the Coulomb and exchange integrals $U_{mm}=U_{0}=0.2749$ Ry, $U_{mm'}=U_{1}=0.1426$ Ry, and $J_{mm'}=J=0.0662$ Ry. These values are obtained from the relations $U_{0}=\bar{U}+8 \bar{J}/5, U_{1}=\bar{U}-2\bar{J}/5,$ and $J=\bar{J}$, using the average values $\bar{U}=0.1691$ Ry and $\bar{J}=0.0662$ Ry by Anisimov {\it{et al.}}\cite{via97}. Note that we adopted here the relation $U_{0}=U_{1}+2J$ for the cubic system. 

\begin{figure}[!htbp]
\begin{center}
\includegraphics[width=8cm,angle=270]{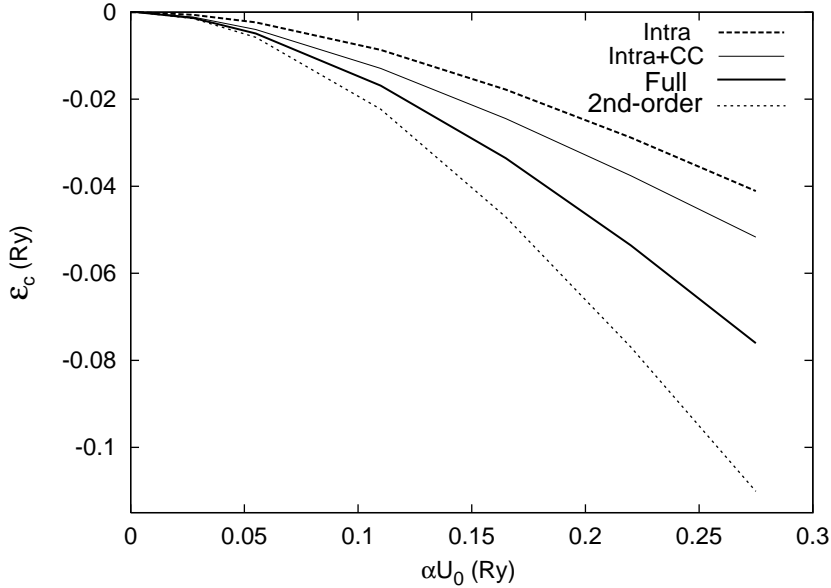}
\end{center}
\vspace{2cm}
\caption{The correlation energy $\epsilon_c$ as a function of Coulomb interaction strength $\alpha U_{0}$. Dashed curve: the result with only the intra-orbital correlations, thin solid curve: the result with both the intra-orbital and inter-orbital charge-charge correlations, solid curve: the result with full correlations. The thin dashed curve indicates the result of the second-order calculations with $\tilde{\eta}_{LL}=\tilde\zeta_{LL'}=1$, and $\tilde{\xi}_{tLL'}=\tilde{\xi}_{lLL'}=-1$. The paramagnetic bcc Fe corresponds to $\alpha U_{0}$=0.27 Ry.}
\label{fig1}
\end{figure}

We solved the self-consistent equations for variational parameters, Eqs. (\ref{equ15}), (\ref{equ33}) $\sim$ (\ref{equ36}), and (\ref{equ40}), and obtained various quantities according to their expressions presented in the last section. In order to understand the systematic change due to the Coulomb interaction strength, we scaled $U_{0}$, $U_{1}$, and $J$ as $\alpha U_{0}$, $\alpha U_{1}$, and $\alpha J$ using a scaling factor $\alpha$. We present the correlation energy $\epsilon_c$ in Fig.\,\ref{fig1} as a function of $\alpha U_{0}$. With increasing $\alpha U_{0}$ (as well as $\alpha U_{1}$ and $\alpha J$), the self-consistent correlation energy $\epsilon_c$ monotonically decreases. The second-order result of $\epsilon_c$ with $\tilde{\eta}_{LL}=\tilde\zeta_{LL'}=1$, and $\tilde{\xi}_{tLL'}=\tilde{\xi}_{lLL'}=-1$ starts to deviate from the self-consistent $\epsilon_c$ at $\alpha U_{0}\approx $ 0.05 Ry, and overestimates the energy gain beyond the value.

In the first-principles MLA, we can describe the intra-orbital, the inter-orbital charge-charge, and the inter-orbital spin-spin correlations by means of the correlators, $\tilde{O}_{iLL}^{(0)}$, $\tilde{O}_{iLL'}^{(1)}$, and $\tilde{O}_{iLL'}^{(2)}$. When we take into account only the intra-orbital correlations, we find the correlation energy $\epsilon_c=-0.041$ Ry for $\alpha U_{0} =0.27$ Ry ($,i.e.,$ for Fe). When we take into account both the intra-orbital and inter-orbital charge-charge correlations, the correlation energy decreases and $\epsilon_c = -0.050$ Ry for Fe. When we add the inter-orbital spin-spin correlations, the correlation energy decreases further and we obtain $\epsilon_c = -0.076$ Ry for Fe. We find that the inter-orbital correlations make a significant contribution to the correlation energy.

\begin{figure}
\begin{center}
\includegraphics[width=8cm,angle=270]{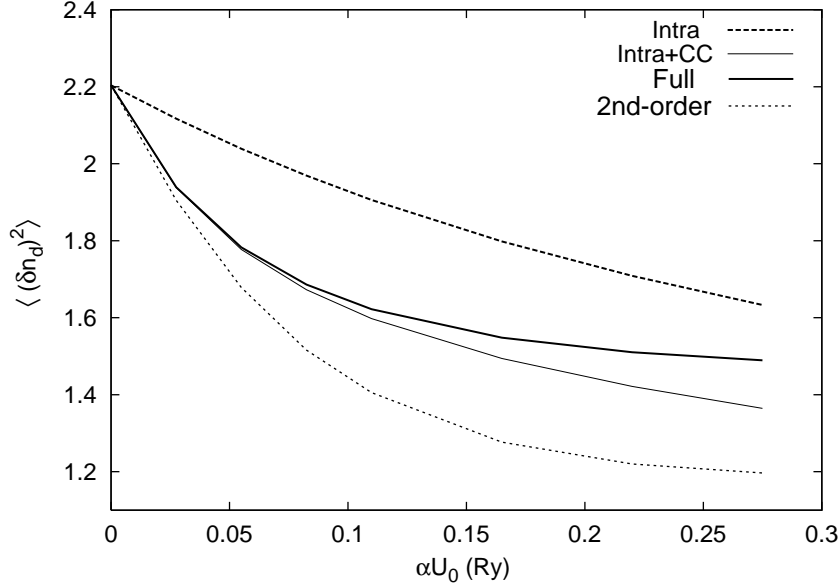}
\end{center}
\vspace{2cm}
\caption{The charge fluctuation $\langle (\delta n_{d})^2\rangle$ vs Coulomb interaction strength $\alpha U_{0}$ curves. Dashed curve: the result with only the intra-orbital correlations, thin solid curve: the result with both the intra-orbital and inter-orbital charge-charge correlations, solid curve: the result with full correlations. The thin dashed curve indicates the second-order result. The paramagnetic bcc Fe corresponds to $\alpha U_{0}$=0.27 Ry.}
\label{fig2}
\end{figure}

The correlation energy gain is accompanied by the suppression of charge fluctuations. We calculated the charge fluctuations for $d$ electrons $\langle (\delta n_{d})^2\rangle=\langle n_{d}^2\rangle- \langle n_{d}\rangle^2$ as a function of $\alpha U_{0}$ as shown in Fig.\,\ref{fig2}. The charge fluctuation in the Hartree-Fock approximation is 2.20. It is suppressed rapidly with increasing the Coulomb interaction strength $\alpha U_{0}$. We obtain the charge fluctuations $\langle (\delta n_{d})^2\rangle$=1.51 for $\alpha U_{0}=0.27$ Ry (Fe). The lowest-order result of calculations deviates downward from the self-consistent result even for a small $\alpha U_{0}$ with increasing $\alpha U_{0}$; it overestimates the suppression of charge fluctuations. We examined the contributions of the three kinds of correlations to $\langle (\delta n_{d})^2\rangle$. The intra-orbital correlations suppress the charge fluctuations, and lead to $\langle (\delta n_{d})^2\rangle=1.73$ for $\alpha U_{0}=0.27$ Ry (Fe). The inter-orbital charge-charge correlations decrease the charge fluctuations further, and we have $\langle (\delta n_{d})^2\rangle$=1.36 for Fe. The result is comparable to the value of the LA with the $d$-band model\cite{pfu95}, $i.e.,$ $\langle (\delta n_{d})^2\rangle \approx 1.0$, though it is somewhat larger than that of the LA because the present theory takes into account the hybridization between the $d$ and $sp$ electrons and the latter delocalizes the $d$ electrons. We also notice that the inter-orbital spin-spin correlations also delocalize the $d$ electrons as shown in Fig.\,\ref{fig2}, so that we finally obtain $\langle (\delta n_{d})^2\rangle$=1.51, which is considerably larger than that was obtained by the LA and the $d$ band model.

\begin{figure}
\begin{center}
\includegraphics[width=8cm,angle=270]{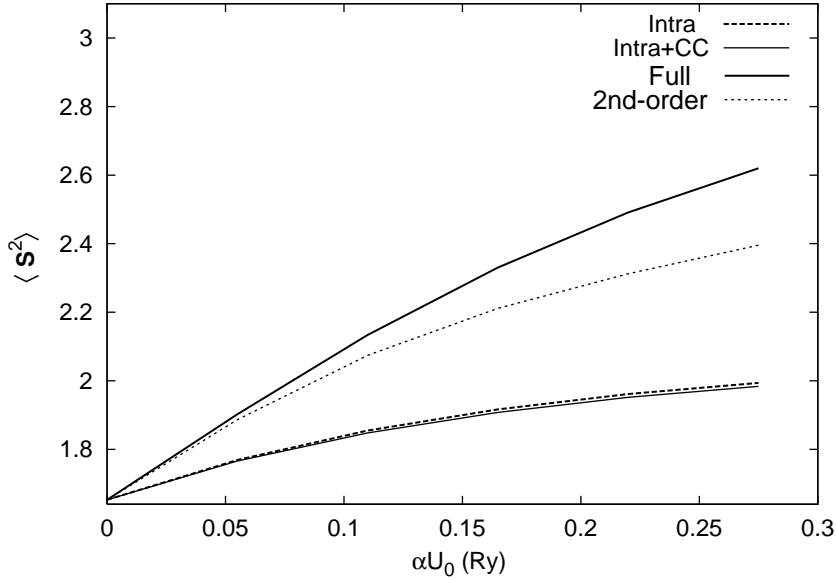}
\end{center}
\vspace{2cm}
\caption{The amplitude of local moment as a function of the Coulomb interaction strength $\alpha U_{0}$. Dashed curve: the result with only the intra-orbital correlations, thin solid curve: the result with both the intra-orbital and inter-orbital charge-charge correlations, solid curve: the result with full correlations, thin dashed curve: the second-order result. The paramagnetic bcc Fe corresponds to $\alpha U_{0}$=0.27 Ry.}
\label{fig3}
\end{figure}

Formation of atomic magnetic moments also originates in the $d$ electron correlations, and determines the magnetic properties of Fe at finite temperatures. We calculated the amplitude of local moment $\langle \bm{S}^2\rangle$ as a function of $\alpha U_{0}$ as shown in Fig.\,\ref{fig3}. We have $\langle \bm{S}^2\rangle=1.65$ for the Hartree-Fock uncorrelated electrons. The amplitudes of local moment monotonically increase with increasing the Coulomb interaction strength $\alpha U_{0}$, and we find $\langle \bm{S}^2\rangle=2.61$ for Fe in the full self-consistent calculations. The lowest-order calculations underestimate the amplitude, and result in $\langle \bm{S}^2\rangle \approx 2.41$ for $\alpha U_{0}=0.27$ Ry (Fe). The self-consistent result is comparable to the value of the LA with the $d$-band model\cite{pfu95}, $\langle \bm{S}^2\rangle \approx 2.91$, but is somewhat smaller than that of the LA because the present theory takes into account the hybridization between the $d$ and $sp$ electrons. It should be noted that the enhancement of amplitude is caused by both the intra-orbital and inter-orbital spin-spin correlations, and the effects of the inter-orbital charge-charge correlations are negligible as seen in Fig.\,\ref{fig3}. Although there are no direct measurements of the amplitude of local moment $\langle \bm{S}^2\rangle$ for the bcc Fe, one can estimate the experimental value from the observed effective Bohr magneton number $p_{\rm{eff}}\,(=3.20)$\cite{mfa44}, because the Rhodes-Wolhfarth ratio of the bcc Fe is equal to 1.0 within $5\%$ error. In this case, we have the experimental value $\langle \bm{S}^2\rangle=p_{\rm{eff}}^{2}/4=2.56$, which is in good agreement with the present result $\langle \bm{S}^2\rangle=2.61$.

\begin{figure}
\begin{center}
\includegraphics[width=8cm,angle=270]{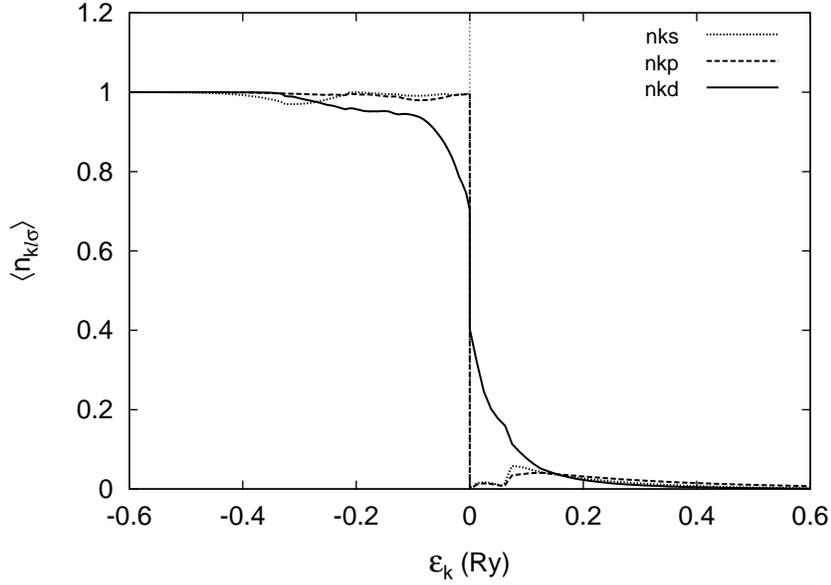}
\end{center}
\vspace{2cm}
\caption{The partial momentum distribution functions $\langle n_{kl\sigma}\rangle$ as a function of the energy $\epsilon_{k}$(=$\epsilon_{kL\sigma}$-$\epsilon_{F}$). Dotted curve: the momentum distribution function for $s$ electrons, dashed curve: the momentum distribution function for $p$ electrons, solid curve: the momentum distribution function for $d$ electrons.}
\label{fig4}
\end{figure}

As we have mentioned in the introduction, the MLA can describe the momentum dependence of the momentum distribution function (MDF). We calculated the partial MDF projected onto each orbital $l$ in order to examine the role of $s$, $p$, and $d$ electrons. They are defined by $\langle n_{kl\sigma}\rangle=\sum_{m}\langle n_{kL\sigma}\rangle/(2l+1)$. Figure \,\ref{fig4} shows the calculated MDF. In the case of $s$ and $p$ electrons the partial MDF are approximately flat below and above the Fermi level $\epsilon_{F}$, and jump at $\epsilon_{F}$. Therefore the $s$ and $p$ electrons behave as independent electrons. The deviation from 1 or 0 are caused by the hybridization with $d$ electrons. On the other hand, the partial MDF for $d$ electrons shows a strong momentum dependence due to electron correlations. 

\begin{figure}
\begin{center}
\includegraphics[width=8cm,angle=270]{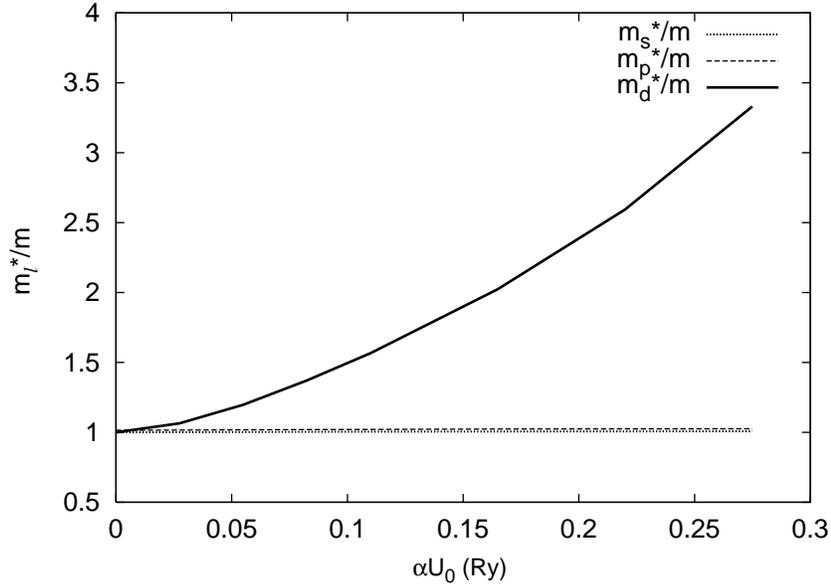}
\end{center}
\vspace{2cm}
\caption{The orbital-dependent mass enhancement $m_{l}^{*}/m$ as a  function of the Coulomb interaction strength $\alpha U_{0}$. Dotted curve: $m_{s}^{*}/m$ ($l$=0), dashed curve: $m_{p}^{*}/m$ ($l$=1), solid curve: $m_{d}^{*}/m$ ($l$=2).}
\label{fig5}
\end{figure}

According to the Fermi liquid theory, the mass enhancement factor ($i.e.,$ the inverse quasiparticle weight) is obtained from the jump at the Fermi level in the MDF. We calculated the orbital-dependent mass enhancement $m_{l}^{*}/m_{0}$ for $s$, $p$ and $d$ electrons as a function of $\alpha U_{0}$ as shown in Fig. \,\ref{fig5}. The $d$ electron mass enhancement rapidly increases with increasing the Coulomb interaction strength $\alpha U_{0}$, while the mass enhancements for the $sp$ electrons almost remain constant and behave as independent electrons irrespective of $\alpha U_{0}$. Calculated mass enhancements are $m_{s}^{*}/m$=$m_{p}^{*}/m$=1.01, and $m_{d}^{*}/m$=3.33 for Fe, respectively. Note that the mass enhancement of the $d$ electrons is significantly larger than the Hartree-Fock value 1.0.

\begin{figure}
\begin{center}
\includegraphics[width=8cm,angle=270]{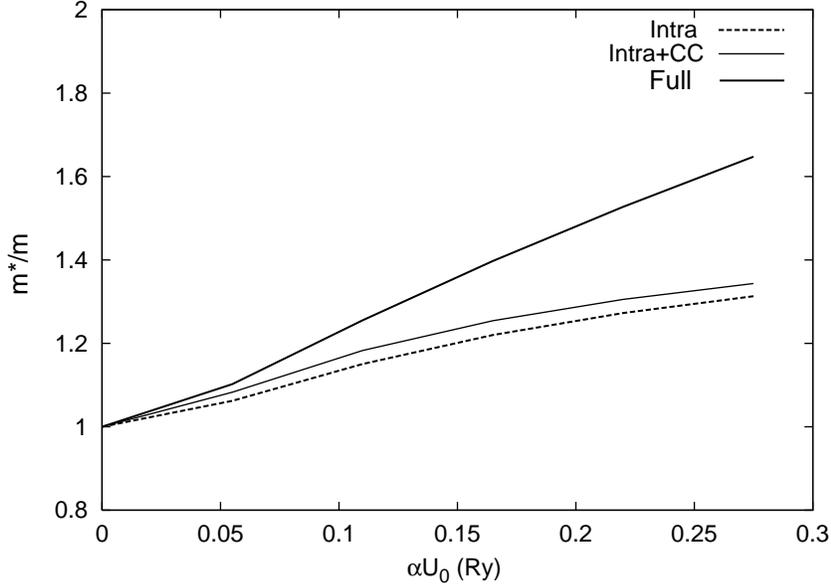}
\end{center}
\vspace{2cm}
\caption{The mass enhancement factor $m^{*}/m$ as a function of the Coulomb interaction strength $\alpha U_{0}$. Dashed curve: mass enhancement due to the intra-orbital correlations, thin solid curve: mass enhancement due to both the intra-orbital and inter-orbital charge-charge correlations, solid curve: the result with the full correlations.}
\label{fig6}
\end{figure}

We calculated the average mass enhancement $m^{*}/m$ $(=1/Z)$ as a function of $\alpha U_{0}$. Calculated $m^{*}/m$ vs Coulomb interaction curve is presented in  Fig.\,\ref{fig6}. The curves with the intra-orbital correlations as well as the curve with both the intra-orbital and inter-orbital charge-charge correlations are also presented there. By comparing these three curves, we find that the mass enhancement $m^{*}/m$ for Fe ($\alpha U_{0}$=0.27 Ry) is dominated by both the intra-orbital and inter-orbital spin-spin correlations, though the inter-orbital charge-charge correlations also make a significant contribution in the weak interaction regime ($\alpha U_{0}\approx$ 0.05 Ry). The mass enhancement factor for Fe is $m^{*}/m$=1.65 in the present calculations. 

The mass enhancement for the bcc Fe has recently been investigated on the basis of the first-principles theories. Katanin $et$ $al.$\cite{aak10} obtained $m^{*}_{t_{2g}}/m$=1.163 for $t_{2g}$ electrons at 1000K with use of the LDA+DMFT combined with the QMC technique, but they could not obtain the mass enhancement for $e_{g}$ electrons because of the non-Fermi liquid behavior due to strong fluctuations in the narrow $e_{g}$ band at finite temperatures. More recently, Pourovski $et$ $al.$\cite{lvp14} performed the LDA+DMFT calculations for bcc Fe with use of the continuous-time QMC technique. They obtained $m^{*}/m$=1.577 at 300 K for bcc Fe being in agreement with our present result $m^{*}/m$=1.65. The first-principles Gutzwiller calculations by Deng $et$ $al.$\cite{xy09} led to a reasonable value $m^{*}/m$=1.56. But they used too large a Coulomb interaction parameter $\bar{U}$=7.0 eV. Recent results based on the LDA+Gutzwiller theory with use of a reasonable value $\bar{U}$=2.5 eV and $\bar{J}$=1.2 eV show that $m^{*}_{e_{g}}/m$=1.08 for $e_{g}$ electrons and $m^{*}_{t_{2g}}/m$=1.05 for $t_{2g}$ electrons,\cite{gbo14} which are too small as compared with the other results of calculations mentioned above. The present result $m^{*}/m$=1.65 is comparable to the experimental value $m^{*}/m=1.38\sim 2.12$ obtained from the low temperature specific heat data\cite{wpe01, chc60, lch03}, and the recent experimental result  $m^{*}/m=1.7$ obtained by the ARPES.\cite{jsb09}

\section{Summary}
We have developed the first-principles MLA on the basis of the tight-binding LDA +U Hamiltonian in order to describe correlated electrons in the real system. The MLA wavefunction is constructed by applying the intra-orbital correlators, the inter-orbital charge-charge correlators, and the inter-orbital spin-spin correlators with momentum-dependent variational parameters to the Hartree-Fock uncorrelated state. The wavefunction reduces to the Rayleigh-Schr\"odinger perturbation theory in the weak Coulomb interaction limit, and describes the ground state of correlated electrons. We derived the self-consistent equations for the variational parameters within the single-site approximation, and obtained the expressions of the physical quantities.

We studied the correlated electron state of the paramagnetic bcc Fe solving the self-consistent equations for momentum-dependent variational parameters. We obtained the correlation energy $\epsilon_{c}=-0.076$ Ry for the paramagnetic Fe, and found that the inter-orbital correlation contribution is comparable to the intra-orbital one in the case of Fe. The charge fluctuations $\langle (\delta n_{d})^2\rangle$ are suppressed with increasing the Coulomb interaction strength. We obtained $\langle (\delta n_{d})^2\rangle$=1.51 for Fe, which is larger than the value $\langle (\delta n_{d})^2\rangle \approx 1.0$ calculated by the LA and the $d$ band model. The discrepancy is partly caused by the hybridization between $sp$ and $d$ electrons and partly caused by the Hund-rule correlations. The amplitude of local moment $\langle \bm{S}^2\rangle$ increases with increasing the Coulomb interaction strength. We obtained $\langle \bm{S}^2\rangle=2.61$ for Fe which is larger than the Hartree-Fock value $\langle \bm{S}^2\rangle=1.65$ because of the Hund-rule correlations, but is somewhat smaller than that of the $d$-band model+LA value $\langle \bm{S}^2\rangle=2.91$ because the present theory takes into account the hybridization between the $sp$ and $d$ electrons. Present result $\langle \bm{S}^2\rangle=2.61$ shows a good agreement with the experimental value 2.56 estimated from the effective Bohr magneton number of the Curie-Weiss susceptibility.

We obtained the expression of the momentum distribution functions (MDF) for Fe. The MDF depends on the momentum $\bm{k}$ via both the energy $\epsilon_{kn\sigma}$ and the eigenvector $u_{Ln}\bm{(k)}$ in the present theory. In order to obtain the physical picture, we examined the MDF projected onto each orbital. From the analyses, we found that the $d$ electrons cause a significant momentum dependence, though the $sp$ electrons behave as independent electrons. We obtained the mass enhancement factors $m_{s}^{*}/m=m_{p}^{*}/m=1.01$, and $m_{d}^{*}/m=3.33$ for $s$, $p$ and $d$ electrons, respectively, indicating that the $d$ electrons behave as correlated electrons. The average mass enhancement $m^{*}/m$ increases with increasing interaction strength. We found that the intra-orbital and inter-orbital spin-spin correlations $, i.e.,$ spin fluctuations cause the mass enhancement of Fe. We obtained $m^{*}/m=1.65$ for Fe. Calculated value 1.65 is consistent with the experimental values obtained from the low-temperature specific heat data $m^{*}/m=1.38\sim 2.12$, and the ARPES data 1.7, as well as the recent theoretical result 1.577 based on the finite-temperature LDA+DMFT. The first-principles Gutzwiller theory underestimates the mass enhancement factor of bcc Fe, indicating the significance of the momentum dependence of the variational parameters in the MLA. 

Needless to say, the present calculations for Fe are limited to the paramagnetic state. We have to perform the ferromagnetic calculations to clarify the ground state of Fe, Co, and Ni on the basis of the first-principles MLA. Furthermore we have to examine the quantitative aspects of the theory for more correlated electron systems such as Fe pnictides and the heavyfermion compounds. 

The present theory is based on the single-site approximation. It is well known that long-range spin fluctuations can cause a large magnon mass enhancement \cite{pfu83}. We have to include nonlocal correlations to describe quantitatively the magnetism, the metal-insulator transition, and the frustrated electron system, especially, in low-dimensional system where nonlocal charge and spin fluctuations become significant. We leave these problems for future work.

\begin{acknowledgment}


The present work is supported by a Grant-in-Aid for Scientific Research (25400404).

\end{acknowledgment}

\appendix

\section{Correlation Energy in the SSA}
 In this Appendix we derive the correlation energy per atom in the single-site approximation (SSA), $i.e.,$ Eq. (\ref{equ15}). 

The correlation energy is given by 
\begin{align}
\langle\tilde{H}\rangle=\frac{{\langle\Psi_{\mathrm{MLA}}\vert\tilde{H}\vert\Psi_{\mathrm{MLA}}\rangle}}{{\langle\Psi_{\mathrm{MLA}}\vert\Psi_{\mathrm{MLA}}\rangle}}=\frac{A_{N}}{B_{N}}\,.
\label{equa1}
\end{align}
Here $A_{N}$ and $B_{N}$ are defined as follows:
\begin{align}
A_{N}={\left\langle\left[\prod_{i}\left(1-\tilde{O}_{i}^{\dagger}\right) \right]\tilde{H}\left[\prod_{i}\left(1-\tilde{O}_{i}\right)\right]\right\rangle_{0}},
\label{equa2}
\end{align}
\begin{align}
B_{N}={\left\langle\left[\prod_{i}\left(1-\tilde{O}_{i}^{\dagger}\right) \right]\left[\prod_{i}\left(1-\tilde{O}_{i}\right)\right]\right\rangle_{0}}.
\label{equa3}
\end{align}

Expanding $B_{N}$ with respect to site 1, we obtain
\begin{align}
B_{N}&=B_{N-1}^{(1)}-{\left\langle \tilde{O}_{1}^{\dagger}\left[{\prod_{i}}^{(1)}\left(1-\tilde{O}_{i}^{\dagger}\right) \right]\left[{\prod_{i}}^{(1)}\left(1-\tilde{O}_{i}\right)\right]\right\rangle_{0}}\nonumber \\
&-{\left\langle \left[{\prod_{i}}^{(1)}\left(1-\tilde{O}_{i}^{\dagger}\right) \right]\tilde{O}_{1}\left[{\prod_{i}}^{(1)}\left(1-\tilde{O}_{i}\right)\right]\right\rangle_{0}}\nonumber \\
&+{\left\langle\tilde{O}_{1}^{\dagger} \left[{\prod_{i}}^{(1)}\left(1-\tilde{O}_{i}^{\dagger}\right) \right]\tilde{O}_{1}\left[{\prod_{i}}^{(1)}\left(1-\tilde{O}_{i}\right)\right]\right\rangle_{0}},
\label{equa4b}
\end{align}
and 
\begin{align}
B_{N-1}^{(1)}={\left\langle\left[{\prod_{i}}^{(1)}\left(1-\tilde{O}_{i}^{\dagger}\right) \right]\left[{\prod_{i}}^{(1)}\left(1-\tilde{O}_{i}\right)\right]\right\rangle_{0}}.
\label{equa5}
\end{align}
Here the product ${\prod_{i}}^{(1)}$ means the product with respect to all sites except site 1. 

When we apply Wick's theorem for the calculations of $B_{N}$, we neglect the contractions between different sites. This is the SSA, and then Eq. (\ref{equa4b}) is expressed as
\begin{align}
B_{N}=\left\langle\left(1-\tilde{O}_{1}^{\dagger}\right)\left(1-\tilde{O}_{1}\right)\right \rangle_{0}B_{N-1}^{(1)}\,.
\label{equa6}
\end{align}
We adopt the same approximation for $A_{N}$. In this case, there are two-types of terms, the terms in which the operator $\tilde{O}_{1}$ is contracted to $\tilde{H}$ and the other terms in which $\tilde{H}$ is contracted to the other operators $\tilde{O}_{i} (i\neq 1)$. We have then in the SSA
\begin{align}
A_{N}=\left\langle\left(1-\tilde{O}_{1}^{\dagger}\right)\tilde{H}\left(1-\tilde{O}_{1}\right) \right\rangle_{0}B_{N-1}^{(1)}+\left\langle\left(1-\tilde{O}_{1}^{\dagger}\right)\left(1-\tilde{O}_{1}\right) \right\rangle_{0}A_{N-1}^{(1)}\,,
\label{equa7}
\end{align}
and 
\begin{align}
A_{N-1}^{(1)}={\left\langle\left[{\prod_{i}}^{(1)}\left(1-\tilde{O}_{i}^{\dagger}\right) \right]\tilde{H}\left[{\prod_{i}}^{(1)}\left(1-\tilde{O}_{i}\right)\right]\right\rangle_{0}}.
\label{equa8}
\end{align}

Successive application of the recursive relations (\ref{equa7}) and (\ref{equa6}) leads to the following expressions.
\begin{align}
A_{N}=\sum_{i}\left\langle\left(1-\tilde{O}_{i}^{\dagger}\right)\tilde{H}\left(1-\tilde{O}_{i}\right) \right\rangle_{0}B_{N-1}^{(i)}\,.
\label{equa9}
\end{align}
\begin{align}
B_{N}=\prod_{i}\left\langle\left(1-\tilde{O}_{i}^{\dagger}\right)\left(1-\tilde{O}_{i}\right) \right\rangle_{0}=\left\langle\left(1-\tilde{O}_{i}^{\dagger}\right)\left(1-\tilde{O}_{i}\right) \right\rangle_{0}B_{N-1}^{(i)}\,.
\label{equa10}
\end{align}
Taking the ratio $A_{N}/B_{N}$, we obtain the correlation energy as follows.
\begin{align}
N\epsilon_{c}=\langle\tilde{H}\rangle=\sum_{i}\frac{\langle(1-\tilde{O}_{i}^{\dagger})\tilde{H}(1-\tilde{O}_{i})\rangle_{0}}{\langle(1-\tilde{O}_{i}^{\dagger})(1-\tilde{O}_{i})\rangle_{0}}\,.
\label{equa11}
\end{align}
Assuming a site per unit cell and using the relation $\langle \tilde{O}_{i}^{\dagger}\rangle_{0}=\langle \tilde{O}_{i}\rangle_{0}=0$, we obtain the correlation energy per site as follows. 
\begin{equation}
{\epsilon_c} =\frac{{- \langle {\tilde{O_i}^\dagger}} \tilde{H}\rangle_0 -\langle \tilde{H} \tilde{O_i}\rangle_0 
+\langle {\tilde{O_i}^\dagger} \tilde{H} \tilde{O_i}\rangle_0 }{1+\langle{\tilde{O_i}^\dagger\tilde{O_i}}\rangle_0}\,. 
\label{equa12}
\end{equation}
Since the Hamiltonian $\tilde{H}$ is expressed by $\tilde{H}=\tilde{H}_{0}+{H}_{I}$ and $\langle\tilde{O_i}^\dagger\tilde{H}_{0}\rangle=0$, we obtain the correlation energy $\epsilon_{c}$ as follows.
\begin{equation}
{\epsilon_c} =\frac{{- \langle {\tilde{O_i}^\dagger}} {H}_{I}\rangle_0 -\langle {H}_{I} \tilde{O_i}\rangle_0 
+\langle {\tilde{O_i}^\dagger} \tilde{H} \tilde{O_i}\rangle_0 }{1+\langle{\tilde{O_i}^\dagger\tilde{O_i}}\rangle_0}\,. 
\label{equa13}
\end{equation}
This is the correlation energy in the SSA given in Eq. (\ref{equ15}) in \S 2.

\section{Self-Consistent Solution of Eq. (\ref{equ22}) in the Weak Coulomb Interaction Limit}
In this Appendix, we present the explicit expressions of $P_{LL'L''L'''}^{(\alpha\alpha')}(\{2'21'1\})$  $Q_{LL'L''L'''}^{(\alpha\alpha')}(\{2'21'1\}\{4'43'3\})$, and $S_{LL'L''L'''}^{(\alpha\alpha')}(\{2'21'1\}\{4'43'3\})$ in the self-consistent equation (\ref{equ22}), and derive the solution (\ref{equ23}) in the weak Coulomb interaction limit. 

In the weak Coulomb interaction limit, the third term at the lhs of the self-consistent equation (\ref{equ22}) can be neglected because it is higher order in $\{U_{LL'}^{(\alpha)}\}$. Equation (\ref{equ22}) is then expressed as follows.
\begin{align}
&\sum_{\alpha'}\sum_{<L''L'''>}\sum_{\{kn\sigma\}}^{4'43'3}\Big[Q_{LL'L''L'''}^{(\alpha\alpha')}(\{2'21'1\}\{4'43'3\})\nonumber \\
&-\epsilon_{c}\ S_{LL'L''L'''}^{(\alpha\alpha')}(\{2'21'1\}\{4'43'3\})\Big]\ \lambda_{L''L'''\{4'43'3\}}^{(\alpha')}
=\sum_{\alpha'}\sum_{<L''L'''>}U_{L''L'''}^{(\alpha')}P_{L''L'''LL'}^{(\alpha'\alpha)*}(\{2'21'1\})\,.
\label{equ1b}
\end{align}
Here $Q_{LL'L''L'''}^{(\alpha\alpha')}(\{2'21'1\}\{4'43'3\})$, $S_{LL'L''L'''}^{(\alpha\alpha')}(\{2'21'1\}\{4'43'3\})$, and $P_{LL'L''L'''}^{(\alpha\alpha')}(\{2'21'1\})$ are obtained with use of Wick's theorem as follows.
\begin{align}
&Q_{LL'L''L'''}^{(\alpha\alpha')}(\{2'21'1\}\{4'43'3\})\nonumber \\
&=\langle iL\vert k'_{2}n'_{2}\rangle_{\sigma'_{2}}\langle k_{2}n_{2}\vert iL\rangle_{\sigma_{2}}
\langle iL'\vert k'_{1}n'_{1}\rangle_{\sigma'_{1}}\langle k_{1}n_{1}\vert iL'\rangle_{\sigma_{1}}\nonumber \\
&\times\langle k'_{4}n'_{4}\vert  iL''\rangle_{\sigma'_{4}}\langle iL''\vert k_{4}n_{4}\rangle_{\sigma_{4}}\langle k'_{3}n'_{3}\vert iL'''\rangle_{\sigma'_{3}}\langle iL'''\vert k_{3}n_{3}\rangle_{\sigma_{3}}\nonumber \\
&\times\Delta E(\{k'_{2}n'_{2}\sigma'_{2}k_{2}n_{2}\sigma_{2}k'_{1}n'_{1}\sigma'_{1}k_{1}n_{1}\sigma_{1}\})\nonumber \\
&\times(\delta_{14}\delta_{23}\delta_{1'4'}\delta_{2'3'}-\delta_{14}\delta_{23}\delta_{1'3'}\delta_{2'4'}+\delta_{13}\delta_{24}\delta_{1'3'}\delta_{2'4'}-\delta_{13}\delta_{24}\delta_{1'4'}\delta_{2'3'})\nonumber \\
&\times\langle n_{k_{1}n_{1}\sigma_{1}}\rangle_{0} (1-\langle n_{k'_{1}n'_{1}\sigma'_{1}}\rangle_{0})\langle n_{k_{2}n_{2}\sigma_{2}}\rangle_{0} (1-\langle n_{k'_{2}n'_{2}\sigma'_{2}}\rangle_{0})\,,
\label{equ2b}
\end{align}
\begin{align}
&S_{LL'L''L'''}^{(\alpha\alpha')}(\{2'21'1\}\{4'43'3\})\nonumber \\
&=\Big(\langle k'_{1}n'_{1}\vert iL''\rangle_{\sigma'_{1}}\langle  iL''\vert k_{1}n_{1}\rangle_{\sigma_{1}}\langle k'_{2}n'_{2}\vert iL'''\rangle_{\sigma'_{2}}\langle  iL'''\vert k_{2}n_{2}\rangle_{\sigma_{2}}
\delta_{14}\delta_{1'4'}\delta_{23}\delta_{2'3'}\nonumber \\
&-\langle k'_{2}n'_{2}\vert iL''\rangle_{\sigma'_{2}}\langle  iL''\vert k_{1}n_{1}\rangle_{\sigma_{1}}
\langle k'_{1}n'_{1}\vert iL'''\rangle_{\sigma'_{1}}\langle  iL'''\vert k_{2}n_{2}\rangle_{\sigma_{2}}
\delta_{14}\delta_{1'3'}\delta_{23}\delta_{2'4'}\nonumber \\
&-\langle k'_{1}n'_{1}\vert iL''\rangle_{\sigma'_{1}}\langle  iL''\vert k_{2}n_{2}\rangle_{\sigma_{2}}
\langle k'_{2}n'_{2}\vert iL'''\rangle_{\sigma'_{2}}\langle  iL'''\vert k_{1}n_{1}\rangle_{\sigma_{1}}
\delta_{13}\delta_{1'4'}\delta_{24}\delta_{2'3'}\nonumber \\
&+\langle k'_{2}n'_{2}\vert iL''\rangle_{\sigma'_{2}}\langle  iL''\vert k_{2}n_{2}\rangle_{\sigma_{2}}
\langle k'_{1}n'_{1}\vert iL'''\rangle_{\sigma'_{1}}\langle  iL'''\vert k_{1}n_{1}\rangle_{\sigma_{1}}
\delta_{13}\delta_{1'3'}\delta_{24}\delta_{2'4'}\Big)\nonumber \\
&\times\langle iL \vert k'_{2}n'_{2} \rangle_{\sigma'_{2}}\langle k_{2}n_{2} \vert iL\rangle_{\sigma_{2}}\langle iL'\vert k'_{1}n'_{1} \rangle_{\sigma'_{1}}\langle k_{1}n_{1}\vert iL'\rangle_{\sigma_{1}}\nonumber \\
&\times\langle n_{k_{1}n_{1}\sigma_{1}}\rangle_{0} (1-\langle n_{k'_{1}n'_{1}\sigma'_{1}}\rangle_{0})\langle n_{k_{2}n_{2}\sigma_{2}}\rangle_{0} (1-\langle n_{k'_{2}n'_{2}\sigma'_{2}}\rangle_{0})\,,
\label{equ3b}
\end{align}
\begin{align}
&P_{LL'L''L'''}^{(\alpha\alpha')}(\{2'21'1\})\nonumber \\
&=\sum_{\{ kn\sigma\}}^{44'33'}C_{\sigma'_{4}\sigma_{4}\sigma'_{3}\sigma_{3}}^{(\alpha)}\langle k'_{4}n'_{4}\vert  iL\rangle_{\sigma'_{4}}\langle iL\vert k_{4}n_{4}\rangle_{\sigma_{4}}\langle k'_{3}n'_{3}\vert iL'\rangle_{\sigma'_{3}}\langle iL'\vert k_{3}n_{3}\rangle_{\sigma_{3}}\nonumber \\
&\times\Big(\delta_{2'4}\delta_{24'}\delta_{1'3}\delta_{13'}-\delta_{2'3}\delta_{24'}\delta_{1'4}\delta_{13'}-\delta_{24'}\delta_{23'}\delta_{1'3}\delta_{14'}+\delta_{2'3}\delta_{23'}\delta_{1'4}\delta_{14'}\Big)\nonumber \\
&\times\langle k'_{2}n'_{2}\vert iL''\rangle_{\sigma'_{2}}\langle iL''\vert k_{2}n_{2} \rangle_{\sigma_{2}}\langle k'_{1}n'_{1}\vert iL'''\rangle_{\sigma'_{1}}\langle iL''' \vert k_{1}n_{1} \rangle_{\sigma_{1}}\nonumber \\
&\times\langle n_{k_{1}n_{1}\sigma_{1}}\rangle_{0} (1-\langle n_{k'_{1}n'_{1}\sigma'_{1}}\rangle_{0})\langle n_{k_{2}n_{2}\sigma_{2}}\rangle_{0} (1-\langle n_{k'_{2}n'_{2}\sigma'_{2}}\rangle_{0})\,.
\label{equ4b}
\end{align}

Substituting Eqs. (\ref{equ2b}) $\sim$ (\ref{equ4b}) into Eq. (\ref{equ1b}), we obtain
\begin{align}
&\sum_{\alpha'}\sum_{<L''L'''>}\Big(\Delta E(\{k'_{2}n'_{2}\sigma'_{2}k_{2}n_{2}\sigma_{2}k'_{1}n'_{1}\sigma'_{1}k_{1}n_{1}\sigma_{1}\})-\epsilon_{c}\Big)\nonumber \\
&\times\Big[\langle k'_{1}n'_{1}\vert iL''\rangle_{\sigma'_{1}}\langle iL''\vert k_{1}n_{1}\rangle_{\sigma_{1}}\langle k'_{2}n'_{2}\vert iL'''\rangle_{\sigma'_{2}}\langle iL'''\vert k_{2}n_{2} \rangle_{\sigma_{2}}\lambda_{L''L'''\{1'1 2'2\}}^{(\alpha')}\nonumber \\
&-\langle k'_{2}n'_{2}\vert iL''\rangle_{\sigma'_{2}}\langle iL''\vert k_{1}n_{1} \rangle_{\sigma_{1}}\langle k'_{1}n'_{1}\vert iL'''\rangle_{\sigma'_{1}}\langle iL'''\vert k_{2}n_{2} \rangle_{\sigma_{2}}\lambda_{L''L'''\{2'1 1'2\}}^{(\alpha')}\nonumber \\
&-\langle k'_{1}n'_{1}\vert iL''\rangle_{\sigma'_{1}}\langle iL''\vert k_{2}n_{2} \rangle_{\sigma_{2}}\langle k'_{2}n'_{2}\vert iL'''\rangle_{\sigma'_{2}}\langle iL'''\vert k_{1}n_{1} \rangle_{\sigma_{1}}\lambda_{L''L'''\{1'2 2'1\}}^{(\alpha')}\nonumber \\
&+\langle k'_{2}n'_{2}\vert iL''\rangle_{\sigma'_{2}}\langle iL''\vert k_{2}n_{2} \rangle_{\sigma_{2}}\langle k'_{1}n'_{1}\vert iL'''\rangle_{\sigma'_{1}}\langle iL'''\vert k_{1}n_{1} \rangle_{\sigma_{1}}\lambda_{L''L'''\{2'2 1'1\}}^{(\alpha')}\Big]\nonumber \\
&=\sum_{\alpha'}\sum_{<L''L'''>}U_{L''L'''}^{(\alpha')}\Big[C_{\sigma_{2}\sigma'_{2}\sigma_{1}\sigma'_{1}}^{(\alpha')}\langle iL''\vert k_{2}n_{2}\rangle_{\sigma_{2}}\langle k'_{2}n'_{2}\vert iL'' \rangle_{\sigma'_{2}}\langle iL'''\vert k_{1}n_{1}\rangle_{\sigma_{1}}\langle k'_{1}n'_{1}\vert iL''' \rangle_{\sigma'_{1}}\nonumber \\
&\hspace{2.9cm}-C_{\sigma_{2}\sigma'_{1}\sigma_{1}\sigma'_{2}}^{(\alpha')}\langle iL''\vert k_{2}n_{2}\rangle_{\sigma_{2}}\langle k'_{1}n'_{1}\vert iL'' \rangle_{\sigma'_{1}}\langle iL'''\vert k_{1}n_{1}\rangle_{\sigma_{1}}\langle k'_{2}n'_{2}\vert iL''' \rangle_{\sigma'_{2}}\nonumber \\
&\hspace{2.9cm}-C_{\sigma_{1}\sigma'_{2}\sigma_{2}\sigma'_{1}}^{(\alpha')}\langle iL''\vert k_{1}n_{1}\rangle_{\sigma_{1}}\langle k'_{2}n'_{2}\vert iL'' \rangle_{\sigma'_{2}}\langle iL'''\vert k_{2}n_{2}\rangle_{\sigma_{2}}\langle k'_{1}n'_{1}\vert iL''' \rangle_{\sigma'_{1}}\nonumber \\
&\hspace{2.9cm}+C_{\sigma_{1}\sigma'_{1}\sigma_{2}\sigma'_{2}}^{(\alpha')}\langle iL''\vert k_{1}n_{1}\rangle_{\sigma_{1}}\langle k'_{1}n'_{1}\vert iL'' \rangle_{\sigma'_{1}}\langle iL'''\vert k_{2}n_{2}\rangle_{\sigma_{2}}\langle k'_{2}n'_{2}\vert iL''' \rangle_{\sigma'_{2}}\Big]\,.
\label{equ5b}
\end{align}
Here $C_{\sigma_{2}\sigma'_{2}\sigma_{1}\sigma'_{1}}^{(\alpha)}$ are defined by Eq. (\ref{equ23ab}). 

Using the expression of the overlap integral $\langle iL\vert kn\rangle_{\sigma}=u_{Ln\sigma}(\bm{k}) e^{-i\bm{k}\cdot \bm{R}_{i}}/\sqrt{N}$ in Eq. (\ref{equ5b}), and defining $a_{LL'\{2'21'1\}}$ by
\begin{equation}
a_{LL'\{2'21'1\}}=u_{Ln'_{2}\sigma'_{2}}^{*}(\bm{k'}_{2})\ u_{Ln_{2}\sigma_{2}}(\bm{k}_{2})\ u_{L' n'_{1}\sigma'_{1}}^{*}(\bm{k'}_{1})\ u_{L' n_{1}\sigma_{1}}(\bm{k}_{1})\, ,
\label{equa19}
\end{equation}
we obtain the self-consistent equation as follows.
\begin{align}
&\sum_{\alpha}\sum_{<LL'>}\left(a_{LL'\{2'21'1\}}\lambda_{LL'\{2'21'1\}}^{(\alpha)}-a_{LL'\{1'22'1\}}\lambda_{LL'\{1'22'1\}}^{(\alpha)}-a_{LL'\{2'11'2\}}\lambda_{LL'\{2'11'2\}}^{(\alpha)}+a_{LL'\{1'12'2\}}\lambda_{LL'\{1'12'2\}}^{(\alpha)}\right)\nonumber \\
&=\sum_{\alpha}\sum_{<LL'>}\Big(\Delta E(\{k'_{2}n'_{2}\sigma'_{2}k_{2}n_{2}\sigma_{2}k'_{1}n'_{1}\sigma'_{1}k_{1}n_{1}\sigma_{1}\})-\epsilon_{c}\Big)^{-1}\nonumber \\
&\times U_{LL'}^{(\alpha)}\left(C_{\sigma_{2}\sigma'_{2}\sigma_{1}\sigma'_{1}}^{(\alpha)} a_{LL'\{2'21'1\}}-C_{\sigma'_{1}\sigma_{2}\sigma'_{2}\sigma'_{1}}^{(\alpha)} a_{LL'\{1'22'1\}}-C_{\sigma_{2}\sigma'_{1}\sigma_{1}\sigma'_{2}}^{(\alpha)} a_{LL'\{2'11'2\}}+C_{\sigma_{1}\sigma'_{1}\sigma_{2}\sigma'_{2}}^{(\alpha)}a_{LL'\{1'12'2\}}\right)\,.
\label{equa20}
\end{align}
Then, we find the following solution by inspection.
\begin{equation}
\lambda^{(\alpha)}_{{LL'}\{{2'2 1'1}\}}=\frac{C_{\sigma_{2}\sigma_{2}^{'}\sigma_{1}\sigma_{1}^{'}}^{(\alpha)}U_{LL'}^{(\alpha)}}{\Delta E_{k'_{2}n'_{2}\sigma'_{2} k_{2}n_{2}\sigma_{2} k'_{1}n'_{1}\sigma'_{1} k_{1}n_{1}\sigma_{1}}-\epsilon_c}\,.
\label{equa23a}
\end{equation}
This is the solution presented in Eq. (\ref{equ23}) in \S 2.

\section{Matrix Elements in the Self-Consistent Equations (\ref{equ33}) $\sim$ (\ref{equ36})} 

In this Appendix we present the expressions of all the matrix elements in the self-consistent Eqs. (\ref{equ33}) $\sim$ (\ref{equ36}). We assume that the orbital $L$ belongs to an irreducible representation $\Gamma$ of the point symmetry with dimensions $d_{\Gamma}$. Moreover we assume for simplicity that the Coulomb interactions $U_{LL'}^{(\alpha)}$ only depend on the types of the irreducible representations $\Gamma$ and $\Gamma'$ to which the orbitals $L$ and $L'$ belong; $U_{LL'}^{(\alpha)}=U_{\Gamma\Gamma'}^{(\alpha)}$. Then the final expressions of the elements for the self-consistent equations (\ref{equ33}) $\sim$ (\ref{equ36}) in the paramagnetic state are given as follows by means of the Laplace transform of the local density of states $\rho_{\Gamma}(\epsilon)$ in the Hartree-Fock approximation. 
\begin{align}
Q_{\ \Gamma\Gamma{'}}&=-\int_{0}^{\infty}{dt\ dt'{e^{i\epsilon_{c} (t+t')}}\Bigl[a_{\Gamma'}(-t-t')\ b_{\Gamma'}(t+t')\ a_{\Gamma}(-t-t')\ b_{\Gamma 1}(t+t')}\nonumber \\
&\hspace{3.3cm}-a_{\Gamma'}(-t-t')\ b_{\Gamma'}(t+t')\ a_{\Gamma 1}(-t-t')\ b_{\Gamma }(t+t')\nonumber \\
&\hspace{3.3cm}+a_{\Gamma'}(-t-t')\ b_{\Gamma' 1}(t+t')\ a_{\Gamma}(-t-t')\ b_{\Gamma }(t+t')\nonumber \\
&\hspace{3.3cm}-a_{\Gamma'1}(-t-t')\ b_{\Gamma'}(t+t')\ a_{\Gamma}(-t-t')\ b_{\Gamma }(t+t')\Bigr]\,.
\label{equb32}
\end{align}
\begin{align}
S_{\Gamma\Gamma'}=-\int_{0}^{\infty}{dt\ dt'{e^{i\epsilon_{c} (t+t')}}\ a_{\Gamma}(-t-t')\ a_{\Gamma'}(-t-t')\ b_{\Gamma}(t+t')\ b_{\Gamma'}(t+t')}\,.
\label{equb33}
\end{align}
\begin{align}
P_{\Gamma\Gamma{'}}=i\int_{0}^{\infty}{dt {e^{i\epsilon_{c}t}}\ a_{\Gamma}(-t)\ a_{\Gamma'}(-t)\ b_{\Gamma}(t)\ b_{\Gamma'}(t)}\,.
\label{equb31}
\end{align}
\begin{align}
K_{\Gamma \Gamma}^{(0)} &= {U}_{\Gamma \Gamma}^{{(0)}^{2}} \Omega_{\Gamma \Gamma}\ \tilde{\lambda}_{0\Gamma \Gamma}+4\ (d_{\Gamma}-1)\ {U}_{\Gamma \Gamma}^{{(1)}^{2}}\ {M}_{\Gamma \Gamma}\ \tilde{\lambda}_{1\Gamma \Gamma}\nonumber \\
& +4\sum_{\Gamma{'}(\neq\Gamma)} d_{\Gamma{'}}\ {U}_{\Gamma \Gamma{'}}^{{(1)}^{2}}\ {M}_{\Gamma \Gamma{'}}\ \tilde{\lambda}_{1\Gamma \Gamma{'}} 
+\frac{1}{4}\ (d_{\Gamma}-1)\ {U}_{\Gamma \Gamma}^{{(2)}^{2}} {M}_{\Gamma \Gamma}\ \Big(\tilde{\lambda}_{2l\Gamma \Gamma}+2\ \tilde{\lambda}_{2t\Gamma \Gamma}\Big)\nonumber \\
&+\frac{1}{4}\sum_{\Gamma{'}(\neq\Gamma)}d_{\Gamma{'}}\ {U}_{\Gamma \Gamma'}^{{(2)}^{2}}{M}_{\Gamma \Gamma'}\ \Big(\tilde{\lambda}_{2l\Gamma \Gamma{'}}+2\ \tilde{\lambda}_{2t\Gamma \Gamma{'}}\Big)\,.
\label{equb14}
\end{align}
\begin{align}
\bar{K}_{\Gamma \Gamma }^{(1)} &= 8\,{U}_{\Gamma \Gamma}^{(1)}\ {U}_{\Gamma \Gamma}^{(0)}\ M_{\Gamma \Gamma}\ \tilde{\lambda}_{0\Gamma \Gamma} \nonumber \\
&+4\ {U}_{\Gamma \Gamma}^{(1)}\ {\left[2\,{U}_{\Gamma \Gamma}^{(0)}\ \Xi_{\Gamma \Gamma\Gamma}+{U}_{\Gamma \Gamma}^{(1)}\ \Omega_{\Gamma \Gamma}+4\,(d_{\Gamma}-2)\ {U}_{\Gamma \Gamma}^{(1)}\ \Xi_{\Gamma \Gamma\Gamma}\right]}\ \tilde{\lambda}_{1\Gamma \Gamma}\nonumber \\
&+16\ {U}_{\Gamma \Gamma}^{(1)}\sum_{\Gamma{''}(\neq\Gamma)}d_{\Gamma{''}}\ {U}_{\Gamma \Gamma{''}}^{(1)}\ \Xi_{\Gamma \Gamma\Gamma{''}}\ \tilde{\lambda}_{1\Gamma \Gamma{''}}-\frac{1}{4}\ {U}_{\Gamma \Gamma}^{{(2)^{2}}}\ \Omega_{\Gamma \Gamma}\ \Big(\tilde{\lambda}_{2l\Gamma \Gamma}+2\ \tilde{\lambda}_{2t\Gamma \Gamma}\Big)\,.
\label{equb15}
\end{align}
\begin{align}
\bar{K}_{\Gamma \Gamma{'} }^{(1)} &=4\ {U}_{\Gamma \Gamma{'}}^{(1)}\ \left({U}_{\Gamma \Gamma}^{(0)}\ M_{\Gamma \Gamma{'}}\ \tilde{\lambda}_{0\Gamma \Gamma}+{U}_{\Gamma{'} \Gamma{'}}^{(0)}\ M_{\Gamma{'} \Gamma}\ \tilde{\lambda}_{0\Gamma{'}\Gamma{'}}\right)\nonumber \\
&+4\ {U}_{\Gamma \Gamma{'}}^{(1)}\ \left({U}_{\Gamma \Gamma}^{(0)}\ \Xi_{\Gamma{'} \Gamma\Gamma} +{U}_{\Gamma{'} \Gamma{'}}^{(0)}\ \Xi_{\Gamma \Gamma{'}\Gamma{'}}+ {U}_{\Gamma \Gamma{'}}^{(1)}\ \Omega_{\Gamma \Gamma{'}}\right)\ \tilde{\lambda}_{1\Gamma \Gamma{'}}\nonumber \\
&+8\ (d_{\Gamma}-1)\ {U}_{\Gamma \Gamma}^{(1)}\ {U}_{\Gamma \Gamma{'}}^{(1)}\ \left(\Xi_{\Gamma \Gamma{'}\Gamma}\ \tilde{\lambda}_{1\Gamma \Gamma}+\Xi_{\Gamma{'} \Gamma\Gamma}\ \tilde{\lambda}_{1\Gamma{'} \Gamma}\right)\nonumber \\
&+8\ (d_{\Gamma{'}}-1)\ {U}_{\Gamma \Gamma{'}}^{(1)}\ {U}_{\Gamma{'} \Gamma{'}}^{(1)}\left(\Xi_{\Gamma \Gamma{'}\Gamma{'}}\ \tilde{\lambda}_{1\Gamma \Gamma{'}}+\Xi_{\Gamma{'} \Gamma\Gamma{'}}\ \tilde{\lambda}_{1\Gamma{'} \Gamma{'}}\right)\nonumber \\
&+8\sum_{\Gamma{''}(\neq\Gamma,\Gamma{'})}d_{\Gamma{''}}\ {U}_{\Gamma \Gamma{''}}^{(1)}\ {U}_{\Gamma{'} \Gamma{''}}^{(1)}\left(\Xi_{\Gamma \Gamma{'}\Gamma{''}}\ \tilde{\lambda}_{1\Gamma \Gamma{''}}+\Xi_{\Gamma{'} \Gamma\Gamma{''}}\ \tilde{\lambda}_{1\Gamma{'} \Gamma{''}}\right)\nonumber \\
&-\frac{1}{4}\ {U}_{\Gamma \Gamma{'}}^{{(2)^{2}}}\ \Omega_{\Gamma \Gamma{'}}\ \Big(\tilde{\lambda}_{2l\Gamma \Gamma{'}}+2\ \tilde{\lambda}_{2t\Gamma \Gamma{'}}\Big)\,.
\label{equb16}
\end{align}
\begin{align}
\bar{K}_{l\Gamma \Gamma }^{(2)}& =\frac{1}{2}\ {U}_{\Gamma \Gamma}^{(2)}\ {U}_{\Gamma \Gamma}^{(0)}\ M_{\Gamma \Gamma}\ \tilde{\lambda}_{0\Gamma \Gamma}-\frac{1}{4}\ {U}_{\Gamma \Gamma}^{(1)}\ {U}_{\Gamma \Gamma}^{(2)}\ \Omega_{\Gamma \Gamma}\ \tilde{\lambda}_{1\Gamma \Gamma}\nonumber \\
&-\frac{1}{4}\ {U}_{\Gamma \Gamma}^{(2)}\ {\left[2\ {U}_{\Gamma \Gamma}^{(0)}\ \Xi_{\Gamma \Gamma\Gamma}-{U}_{\Gamma \Gamma}^{(1)}\ \Omega_{\Gamma \Gamma}-(d_{\Gamma}-2)\ {U}_{\Gamma \Gamma}^{(2)}\ \Xi_{\Gamma \Gamma\Gamma}\right]}\ \tilde{\lambda}_{2l\Gamma \Gamma}\nonumber \\
&+\frac{1}{4}\sum_{\Gamma{''}(\neq\Gamma)}d_{\Gamma{''}}\ {U}_{\Gamma \Gamma}^{{(2)}^{2}}\ \Xi_{\Gamma \Gamma\Gamma{''}}\ \tilde{\lambda}_{2l\Gamma \Gamma{''}}-\frac{1}{8}\ {U}_{\Gamma \Gamma}^{{(2)}^{2}}\ W_{\Gamma \Gamma}\ \tilde{\lambda}_{2t\Gamma \Gamma{''}}\,.
\label{equb17}
\end{align}
\begin{align}
\bar{K}_{l\Gamma \Gamma{'} }^{(2)} &=\frac{1}{4}\ {U}_{\Gamma \Gamma{'}}^{(2)}\left({U}_{\Gamma \Gamma}^{(0)}\ M_{\Gamma \Gamma{'}}\ \tilde{\lambda}_{0\Gamma \Gamma}+{U}_{\Gamma{'} \Gamma{'}}^{(0)}\ M_{\Gamma{'} \Gamma}\ \tilde{\lambda}_{0\Gamma{'} \Gamma{'}}\right)\nonumber \\
&+\frac{1}{4}\ {U}_{\Gamma \Gamma{'}}^{(2)}\left({U}_{\Gamma \Gamma}^{(0)}\ \Xi_{\Gamma{'} \Gamma\Gamma} +{U}_{\Gamma{'} \Gamma{'}}^{(0)}\ \Xi_{\Gamma \Gamma{'}\Gamma{'}}-{U}_{\Gamma \Gamma{'}}^{(1)}\ \Omega_{\Gamma \Gamma{'}}\right)\ \tilde{\lambda}_{2l\Gamma \Gamma{'}}\nonumber \\
&+\frac{1}{8}\ (d_{\Gamma}-1)\ {U}_{\Gamma \Gamma}^{(2)}\ {U}_{\Gamma \Gamma{'}}^{(2)}\left(\Xi_{\Gamma \Gamma{'}\Gamma}\ \tilde{\lambda}_{2l\Gamma \Gamma}+\Xi_{\Gamma{'} \Gamma\Gamma}\ \tilde{\lambda}_{2l\Gamma{'} \Gamma}\right)\nonumber \\
&+\frac{1}{8}\ (d_{\Gamma{'}}-1)\ {U}_{\Gamma \Gamma{'}}^{(2)}\ {U}_{\Gamma{'} \Gamma{'}}^{(2)}\left(\Xi_{\Gamma \Gamma{'}\Gamma{'}}\ \tilde{\lambda}_{2l\Gamma \Gamma{'}}+\Xi_{\Gamma{'} \Gamma\Gamma{'}}\ \tilde{\lambda}_{2l\Gamma{'} \Gamma{'}}\right)\nonumber \\
&-\frac{1}{8}\sum_{\Gamma{''}(\neq\Gamma,\Gamma{'})}d_{\Gamma{''}}\ {U}_{\Gamma{'} \Gamma{''}}^{(2)}\ {U}_{\Gamma \Gamma{''}}^{(2)}\ \left(\Xi_{\Gamma \Gamma{'}\Gamma{''}}\ \tilde{\lambda}_{2l\Gamma \Gamma{''}}+\Xi_{\Gamma{'} \Gamma\Gamma{''}}\ \tilde{\lambda}_{2l\Gamma{'} \Gamma{''}}\right)\nonumber \\
&-\frac{1}{4}\ {U}_{\Gamma \Gamma{'}}^{(1)}\ {U}_{\Gamma \Gamma{'}}^{(2)}\ \Omega_{\Gamma \Gamma{'}}\ \tilde{\lambda}_{1\Gamma \Gamma{'}}-\frac{1}{8}\ {U}_{\Gamma \Gamma{'}}^{{(2)}^{2}}\  W_{\Gamma \Gamma{'}}\ \tilde{\lambda}_{2t\Gamma \Gamma{'}}\,.
\label{equb18}
\end{align}
\begin{align}
K_{t\Gamma \Gamma}^{(2)}&=\frac{1}{2}\ {U}_{\Gamma \Gamma}^{(2)}\ {U}_{\Gamma \Gamma}^{(0)}\ M_{\Gamma \Gamma}\ \tilde{\lambda}_{0\Gamma \Gamma}-\frac{1}{4}\ {U}_{\Gamma \Gamma}^{(2)}\ {U}_{\Gamma \Gamma}^{(1)}\ \Omega_{\Gamma \Gamma}\ \tilde{\lambda}_{1\Gamma \Gamma}\nonumber \\
&-\frac{1}{16}\ {U}_{\Gamma \Gamma}^{{(2)}^{2}}\ W_{\Gamma \Gamma}\ \tilde{\lambda}_{2l\Gamma \Gamma}+\frac{1}{4}\ {U}_{\Gamma \Gamma}^{(2)}\ \Big({U}_{\Gamma \Gamma}^{(1)}+\frac{1}{4}\ {U}_{\Gamma \Gamma}^{(2)}\Big)\ \Omega_{\Gamma \Gamma}\ \tilde{\lambda}_{2t\Gamma \Gamma}\nonumber \\
&+\frac{1}{4}\ (d_{\Gamma}-2)\ {U}_{\Gamma \Gamma}^{{(2)}^{2}}\ \Xi_{\Gamma \Gamma\Gamma}\ \tilde{\lambda}_{2t\Gamma \Gamma}+\frac{1}{4}\sum_{\Gamma{''}(\neq\Gamma)}d_{\Gamma{''}}\ {U}_{\Gamma \Gamma{''}}^{{(2)}^{2}}\ \Xi_{\Gamma \Gamma\Gamma{''}}\ \tilde{\lambda}_{2t\Gamma \Gamma{''}}\,.
\label{equb19}
\end{align}
\begin{align}
K_{t\Gamma \Gamma{'}}^{(2)} &=\frac{1}{4}\ {U}_{\Gamma \Gamma{'}}^{(2)}\left({U}_{\Gamma \Gamma}^{(0)}\ M_{\Gamma \Gamma{'}}\ \tilde{\lambda}_{0\Gamma \Gamma}+{U}_{\Gamma{'} \Gamma{'}}^{(0)}\ M_{\Gamma{'} \Gamma}\ \tilde{\lambda}_{0\Gamma{'} \Gamma{'}}\right)\nonumber \\
&-\frac{1}{4}\ {U}_{\Gamma \Gamma{'}}^{(1)}\ {U}_{\Gamma \Gamma{'}}^{(2)}\ \Omega_{\Gamma \Gamma{'}}\ \tilde{\lambda}_{1\Gamma \Gamma{'}}-\frac{1}{16}\ {U}_{\Gamma \Gamma'}^{{(2)}^{2}}\ W_{\Gamma \Gamma{'}}\ \tilde{\lambda}_{2l\Gamma \Gamma{'}}\nonumber \\
&+\frac{1}{4}\ {U}_{\Gamma \Gamma{'}}^{(2)}\ \Big({U}_{\Gamma \Gamma{'}}^{(1)}+\frac{1}{4}\ {U}_{\Gamma \Gamma{'}}^{(2)}\Big)\ \Omega_{\Gamma \Gamma{'}}\ \tilde{\lambda}_{2t\Gamma \Gamma{'}}\nonumber \\
&+\frac{1}{8}\ (d_{\Gamma}-1)\ {U}_{\Gamma \Gamma}^{(2)}\ {U}_{\Gamma{'} \Gamma}^{(2)}\left(\Xi_{\Gamma \Gamma{'}\Gamma}\ \tilde{\lambda}_{2t\Gamma \Gamma}+\Xi_{\Gamma{'} \Gamma\Gamma}\ \tilde{\lambda}_{2t\Gamma{'} \Gamma}\right)\nonumber \\
&+\frac{1}{8}\ (d_{\Gamma{'}}-1)\ {U}_{\Gamma \Gamma'}^{(2)}\ {U}_{\Gamma' \Gamma'}^{(2)}\left(\Xi_{\Gamma \Gamma{'}\Gamma{'}}\ \tilde{\lambda}_{2t\Gamma \Gamma{'}}+\Xi_{\Gamma{'} \Gamma\Gamma{'}}\ \tilde{\lambda}_{2t\Gamma{'} \Gamma{'}}\right)\nonumber \\
&+\frac{1}{8}\ \sum_{\Gamma{''}(\neq\Gamma,\Gamma{'})}d_{\Gamma{''}}\ {U}_{\Gamma \Gamma{''}}^{(2)}\ {U}_{\Gamma{'} \Gamma{''}}^{(2)}\left(\Xi_{\Gamma \Gamma{'}\Gamma{''}}\ \tilde{\lambda}_{2t\Gamma \Gamma{''}}+\Xi_{\Gamma{'} \Gamma\Gamma{''}}\ \tilde{\lambda}_{2t\Gamma{'} \Gamma{''}}\right)\,.
\label{equb20}
\end{align}
Each element at the rhs of the above expressions (\ref{equb14}) $\sim$ (\ref{equb20}) is expressed  as follows.
\begin{align}
M_{\Gamma\Gamma{'}}=\Xi_{\Gamma\Gamma\Gamma{'}}=-\int_{0}^{\infty}{dt\ dt'{e^{i\epsilon_{c} (t+t')}}\ a_{\Gamma}(-t)\ b_{\Gamma}(t)\ a_{\Gamma}(-t-t')\ b_{\Gamma}(t+t')\ a_{\Gamma{'}}(-t')\ b_{\Gamma'}(t')}\,.
\label{equb34}
\end{align}
\begin{align}
\Xi_{\Gamma\Gamma{'}\Gamma{''}}=-\int_{0}^{\infty}{dt\ dt'{e^{i\epsilon_{c} (t+t')}}\ a_{\Gamma{'}}(-t)\ b_{\Gamma{'}}(t)\ a_{\Gamma}(-t-t')\ b_{\Gamma}(t+t')\ a_{\Gamma{''}}(-t')\ b_{\Gamma{''}}(t')}\,.
\label{equb36}
\end{align}
\begin{align}
\Omega_{\Gamma\Gamma{'}}=-\left(Z_{1\Gamma\Gamma{'}}+Z_{2\Gamma\Gamma{'}}-Z_{3\Gamma\Gamma{'}}-Z_{3\Gamma\Gamma{'}}\right)\,.
\label{equb42}
\end{align}
\begin{align}
W_{\Gamma\Gamma{'}}=\left(Z_{1\Gamma\Gamma{'}}+Z_{2\Gamma\Gamma{'}}+Z_{3\Gamma\Gamma{'}}+Z_{3\Gamma\Gamma{'}}\right)\,.
\label{equb43}
\end{align}
\begin{align}
Z_{1\Gamma\Gamma{'}}=-\int_{0}^{\infty}{dt\ dt'{e^{i\epsilon_{c} (t+t')}}\ a_{\Gamma{'}}(-t)\ b_{\Gamma{'}}(t+t')\ a_{\Gamma}(-t-t')\ b_{\Gamma}(t)\ a_{\Gamma{'}}(-t')\ b_{\Gamma}(t')}\,.
\label{equb37}
\end{align}
\begin{align}
Z_{2\Gamma\Gamma{'}}=-\int_{0}^{\infty}{dt\ dt'{e^{i\epsilon_{c} (t+t')}}\ a_{\Gamma{'}}(-t-t')\ b_{\Gamma{'}}(t)\ a_{\Gamma}(-t)\ b_{\Gamma}(t+t')\ b_{\Gamma{'}}(t')\ a_{\Gamma}(-t')}\,.
\label{equb38}
\end{align}
\begin{align}
Z_{3\Gamma\Gamma{'}}=-\int_{0}^{\infty}{dt\ dt'{e^{i\epsilon_{c} (t+t')}}\ a_{\Gamma{'}}(-t)\ b_{\Gamma{'}}(t+t')\ a_{\Gamma}(-t)\ b_{\Gamma}(t+t')\ a_{\Gamma{'}}(-t')\ a_{\Gamma}(-t')}\,.
\label{equb39}
\end{align}
\begin{align}
Z_{4\Gamma\Gamma{'}}=-\int_{0}^{\infty}{dt\ dt'{e^{i\epsilon_{c} (t+t')}}\ a_{\Gamma{'}}(-t-t')\ b_{\Gamma{'}}(t)\ a_{\Gamma}(-t-t')\ b_{\Gamma}(t)\ b_{\Gamma{'}}(t')\ b_{\Gamma}(t')}\,.
\label{equb40}
\end{align}
Here
\begin{align}
a_{\Gamma}(t)=\int_{-\infty}^{\infty}{d\epsilon \ {e^{-i\epsilon t}}f(\tilde{\epsilon})\ {\rho}_{\Gamma}(\epsilon)}\,,
\label{equb44}
\end{align}
\begin{align}
b_{\Gamma}(t)=\int_{-\infty}^{\infty}{d\epsilon \ {e^{-i\epsilon t}}f(-\tilde{\epsilon})\ {\rho}_{\Gamma}(\epsilon)}\,,
\label{equb45}
\end{align}
\begin{align}
a_{\Gamma 1}(t)=\int_{-\infty}^{\infty}{d\epsilon \ {e^{-i\epsilon t}}\epsilon \ f(\tilde{\epsilon})\ {\rho}_{\Gamma}(\epsilon)}\,,
\label{equb46}
\end{align}
\begin{align}
b_{\Gamma1}(t)=\int_{-\infty}^{\infty}{d\epsilon\ {e^{-i\epsilon t}} \epsilon \ f(-\tilde{\epsilon})\ {\rho}_{\Gamma}(\epsilon)}\,.
\label{equb47}
\end{align}
$\rho_{\Gamma}(\epsilon)$ at the rhs is the local density of states per spin for orbital $L$ belonging to the representation $\Gamma$.
\begin{align}
\rho_{\Gamma}(\epsilon)=\sum_{kn}\vert\langle iL\vert kn\rangle\vert^{2}\ \delta{(\epsilon-\epsilon_{kn})}\,.
\label{equb49}
\end{align}

\section{Correlation Correction to Electron Number} 
In this Appendix, we present the explicit expression of the correlation correction to the electron number of orbital $L$. As shown Eq. (\ref{equ41}), it is given by
\begin{align}
\langle \tilde{n}_{iL}\rangle=\frac{\langle\tilde{O}_{i}^{\dagger}\tilde{n}_{iL}\tilde{O}_{i}\rangle_{0}}{1+\langle{\tilde{O_i}^\dagger\tilde{O_i}}\rangle_0}\,.
\label{equb41a}
\end{align}
The denominator is the renormalization factor of the wavefunction. Expression of $\langle{\tilde{O_i}^\dagger\tilde{O_i}}\rangle_0$ has been given by Eqs. (\ref{equ31}) and (\ref{equb33}). The numerator is expressed as follows with use of the irreducible representation $\Gamma$ with dimensions $d_{\Gamma}$ to which the orbital $L$ belongs.
\begin{align}
\langle\tilde{O}_{i}^{\dagger}\tilde{n}_{iL}\tilde{O}_{i}\rangle_{0}&=2\ A_{\Gamma\Gamma}\ \Bigl[{U}_{\Gamma \Gamma}^{{(0)}^{2}}\ \tilde{\lambda}_{0\Gamma \Gamma}^{2}\nonumber \\
&+(d_{\Gamma}-1)\ A_{\Gamma \Gamma}\ \Big( 2\ {U}_{\Gamma \Gamma}^{{(1)}^{2}}\ \tilde{\lambda}_{1\Gamma \Gamma}^{2}+\frac{1}{8}\ {U}_{\Gamma \Gamma}^{{(2)}^{2}}\ (\tilde{\lambda}_{2l\Gamma \Gamma}^{2}+2\ \tilde{\lambda}_{2t\Gamma \Gamma}^{2})\Big)\Bigr]\nonumber \\
&+2\ \sum_{\Gamma'\neq\Gamma }^{d}d_{\Gamma'}\ A_{\Gamma' \Gamma}\ \Bigl[2\ {U}_{\Gamma \Gamma'}^{{(1)}^{2}}\ \tilde{\lambda}_{1\Gamma \Gamma'}^{2}+\frac{1}{8}\ {U}_{\Gamma \Gamma'}^{{(2)}^{2}}\Big(\tilde{\lambda}_{2l\Gamma \Gamma'}^{2}+2\ \tilde{\lambda}_{2t\Gamma \Gamma'}^{2}\Big)\Bigr]\,.
\label{equb30}
\end{align}
Here
\begin{align}
A_{\Gamma\Gamma{'}}&=-\int_{0}^{\infty}{dt\ dt'{e^{i\epsilon_{c}(t+t')}}\Bigl[a_{\Gamma'}(-t-t')\ b_{\Gamma'}(t+t')\ a_{\Gamma}(-t-t')\ b_{\Gamma }(t)\ b_{\Gamma }(t')}\nonumber \\
&\hspace{3.2cm}-a_{\Gamma'}(-t-t')\ b_{\Gamma'}(t+t')\ b_{\Gamma }(t+t')\ a_{\Gamma }(-t)\ a_{\Gamma }(-t')\Bigr]\,.
\label{equb41}
\end{align}
The functions $a_{\Gamma}(t)$ and $b_{\Gamma}(t)$ have been given in Eqs. (\ref{equb44}) and (\ref{equb45}).

\section{Expressions of the Average Residual Interaction Elements} 
The residual interaction elements $\sum_{<LL'>}\langle O_{iLL'}^{(\alpha)}\rangle$ for $\alpha$= 0, 1, and 2 are given by Eq. (\ref{equ45}):
\begin{equation}
\sum_{<LL'>}\langle O_{iLL'}^{(\alpha)}\rangle =\frac{\displaystyle{- \sum_{<LL'>}\langle {\tilde{O_i}^\dagger}} {O}_{iLL'}^{(\alpha)} \rangle_0- \sum_{<LL'>}\langle {O}_{iLL'}^{(\alpha)} \tilde{O_i}\rangle_0 
+\sum_{<LL'>}\langle {\tilde{O_i}^\dagger} {O}_{iLL'}^{(\alpha)} \tilde{O_i}\rangle_0 }{1+\langle{\tilde{O_i}^\dagger\tilde{O_i}}\rangle_0}\,. 
\label{equ45d}
\end{equation}
In this Appendix, we present the explicit expressions of each term in the numerator of Eq. (\ref{equ45}), which were obtained by using Wick's theorem.
\begin{equation}
\sum_{L}\langle {O}_{iLL}^{(0)}\tilde{O}_{i}\rangle_{0}=\sum_{\Gamma} d_{\Gamma}\ {U}_{\Gamma \Gamma}^{{(0)}}\ {P}_{\Gamma \Gamma}\ \tilde{\lambda}_{0\Gamma \Gamma}\,.
\label{equb2}
\end{equation}
\begin{equation}
\sum_{(L,L')}\langle {O}_{iLL'}^{(1)}\tilde{O}_{i}\rangle_{0}=2\sum_{\Gamma} d_{\Gamma}\  (d_{\Gamma}-1)\  {U}_{\Gamma \Gamma}^{{(1)}}\  {P}_{\Gamma \Gamma}\ \tilde{\lambda}_{1\Gamma \Gamma}+4\sum_{(\Gamma,\Gamma{'})} d_{\Gamma}\  d_{\Gamma{'}} {U}_{\Gamma \Gamma{'}}^{{(1)}}\  {P}_{\Gamma \Gamma{'}}\ \tilde{\lambda}_{1\Gamma \Gamma{'}}\,.
\label{equb3}
\end{equation}
\begin{align}
\sum_{(L,L')}\langle {O}_{iLL'}^{(2)}\tilde{O}_{i}\rangle_{0}&=-\frac{1}{8}\sum_{\Gamma} d_{\Gamma}\  (d_{\Gamma}-1)\  {U}_{\Gamma \Gamma}^{{(2)}}\  {P}_{\Gamma \Gamma}\ \Big(\tilde{\lambda}_{2l\Gamma \Gamma}+2\ \tilde{\lambda}_{2t\Gamma \Gamma}\Big)\nonumber \\
&-\frac{1}{4}\ \sum_{(\Gamma,\Gamma{'})} d_{\Gamma}\  d_{\Gamma{'}} {U}_{\Gamma \Gamma{'}}^{{(2)}}\  {P}_{\Gamma \Gamma{'}}\ \Big(\tilde{\lambda}_{2l\Gamma \Gamma{'}}+2\ \tilde{\lambda}_{2t\Gamma \Gamma'}\Big)\,.
\label{equb4}
\end{align}
\begin{align}
\sum_{L}{\langle \tilde{O}_{i}^{\dagger}O_{iLL}^{(0)}\tilde{O}_{i}\rangle_{0}}&=\sum_{\Gamma}d_{\Gamma}\ {U}_{\Gamma \Gamma}^{{(0)}^{2}}\ \tilde{\lambda}_{0\Gamma \Gamma}^{2}\ \Omega_{\Gamma \Gamma}\nonumber \\
&+\sum_{\Gamma}d_{\Gamma}\ (d_{\Gamma}-1)\ \Big(4\ {U}_{\Gamma \Gamma}^{{(1)}^{2}}\ \tilde{\lambda}_{1\Gamma \Gamma}^{2} -\frac{1}{4}\ {U}_{\Gamma \Gamma}^{{(2)}^{2}}\ \tilde{\lambda}_{2l\Gamma \Gamma}^{2}\Big)\ \Xi_{\Gamma \Gamma\Gamma}\nonumber \\
&+\sum_{(\Gamma,\Gamma' )}d_{\Gamma}\ d_{\Gamma'}\ \Big(4\ {U}_{\Gamma \Gamma'}^{{(1)}^{2}}\ \tilde{\lambda}_{1\Gamma \Gamma'}^{2} -\frac{1}{4}\ {U}_{\Gamma \Gamma'}^{{(2)}^{2}}\ \tilde{\lambda}_{2l\Gamma \Gamma'}^{2}\Big)\ \Big(\Xi_{\Gamma' \Gamma\Gamma}+\Xi_{\Gamma \Gamma'\Gamma'}\Big)\,.
\label{equb21}
\end{align}
\begin{align}
\sum_{(L,L')}{\langle \tilde{O}_{i}^{\dagger}O_{iLL'}^{(1)}\tilde{O}_{i}\rangle_{0}}&=8\sum_{\Gamma}d_{\Gamma}\ (d_{\Gamma}-1)\ {U}_{\Gamma \Gamma}^{(1)}\ \tilde{\lambda}_{1\Gamma \Gamma}\ {U}_{\Gamma \Gamma}^{(0)}\ \tilde{\lambda}_{0\Gamma \Gamma}\  M_{\Gamma \Gamma}\nonumber \\
&+8\sum_{(\Gamma,\Gamma')}d_{\Gamma}\ d_{\Gamma'}\ {U}_{\Gamma \Gamma'}^{(1)}\ \tilde{\lambda}_{1\Gamma \Gamma'}\ \Big({U}_{\Gamma \Gamma}^{(0)}\ \tilde{\lambda}_{0\Gamma \Gamma}\  M_{\Gamma \Gamma'}+{U}_{\Gamma' \Gamma'}^{(0)}\ \tilde{\lambda}_{0\Gamma' \Gamma'}\  M_{\Gamma' \Gamma}\Big)\nonumber \\
&+2\sum_{\Gamma}d_{\Gamma}\ (d_{\Gamma}-1)\ {U}_{\Gamma \Gamma}^{(1)}\ \tilde{\lambda}_{1\Gamma \Gamma}\ \Big({U}_{\Gamma \Gamma}^{(1)}\ \tilde{\lambda}_{1\Gamma \Gamma}\ \Omega_{\Gamma \Gamma}+T_{\Gamma \Gamma}^{(11)}\Big)\nonumber \\
&+4\sum_{(\Gamma,\Gamma')}d_{\Gamma}\ d_{\Gamma'}\ {U}_{\Gamma \Gamma'}^{(1)}\ \tilde{\lambda}_{1\Gamma \Gamma'}\ \Big({U}_{\Gamma \Gamma'}^{(1)}\ \tilde{\lambda}_{1\Gamma \Gamma'}\ \Omega_{\Gamma \Gamma'}+T_{\Gamma \Gamma'}^{(11)}\Big)\nonumber \\
&+\frac{1}{8}\sum_{\Gamma}d_{\Gamma}\ (d_{\Gamma}-1)\ {U}_{\Gamma \Gamma}^{{(2)}^{2}}\ \Big(\tilde{\lambda}_{2l\Gamma \Gamma}^{2}+2\ \tilde{\lambda}_{2t\Gamma \Gamma}^{2}\Big)\ \Omega_{\Gamma \Gamma}\nonumber \\
&+\frac{1}{4}\sum_{(\Gamma,\Gamma')}d_{\Gamma}\ d_{\Gamma'}\ {U}_{\Gamma \Gamma'}^{{(2)}^{2}}\ \Big(\tilde{\lambda}_{2l\Gamma \Gamma'}^{2}+2\ \tilde{\lambda}_{2t\Gamma \Gamma'}^{2}\Big)\ \Omega_{\Gamma \Gamma'}\,.
\label{equb22}
\end{align}
Here
\begin{align}
T_{\Gamma \Gamma'}^{(11)}&=-2\ \Big({U}_{\Gamma \Gamma}^{(1)}\ \Xi_{\Gamma \Gamma'\Gamma}\ \tilde{\lambda}_{1\Gamma \Gamma}+{U}_{\Gamma' \Gamma}^{(1)}\ \Xi_{\Gamma' \Gamma\Gamma}\ \tilde{\lambda}_{1\Gamma' \Gamma}\Big)\nonumber \\
&-2\ \Big({U}_{\Gamma \Gamma'}^{(1)}\ \Xi_{\Gamma \Gamma'\Gamma'}\ \tilde{\lambda}_{1\Gamma \Gamma'}+{U}_{\Gamma' \Gamma'}^{(1)}\ \Xi_{\Gamma' \Gamma\Gamma'}\ \tilde{\lambda}_{1\Gamma' \Gamma'}\Big)\nonumber \\
&+2\sum_{\Gamma''}d_{\Gamma''}\ \Big({U}_{\Gamma \Gamma''}^{(1)}\ \Xi_{\Gamma \Gamma'\Gamma''}\ \tilde{\lambda}_{1\Gamma \Gamma''}+{U}_{\Gamma' \Gamma''}^{(1)}\ \Xi_{\Gamma' \Gamma\Gamma''}\ \tilde{\lambda}_{1\Gamma' \Gamma''}\Big)\,.
\label{equb23}
\end{align}
Finally we have
\begin{align}
\sum_{(L,L')}{\langle \tilde{O}_{i}^{\dagger}O_{iLL'}^{(2)}\tilde{O}_{i}\rangle_{0}}&=\sum_{\Gamma}d_{\Gamma}\ {U}_{\Gamma \Gamma}^{(0)}\ \tilde{\lambda}_{0\Gamma \Gamma}\  \hat{K}_{\Gamma \Gamma}^{(0)}\nonumber \\
&+\frac{1}{2}\sum_{\Gamma}d_{\Gamma}\ (d_{\Gamma}-1)\ \Big[4\ {U}_{\Gamma \Gamma}^{(1)}\ \tilde{\lambda}_{1\Gamma \Gamma}\ \hat{K}_{\Gamma \Gamma1}^{(1)}-\frac{1}{4}\ {U}_{\Gamma \Gamma}^{(2)}\ \tilde{\lambda}_{2l\Gamma \Gamma}\ \hat{K}_{\Gamma \Gamma 2}^{(1)}\Big]\nonumber \\
&+\sum_{(\Gamma,\Gamma')}d_{\Gamma}\ d_{\Gamma'}\ \Big[4\ {U}_{\Gamma \Gamma'}^{(1)}\ \tilde{\lambda}_{1\Gamma \Gamma'}\ \hat{K}_{\Gamma \Gamma'1}^{(1)}-\frac{1}{4}\ {U}_{\Gamma \Gamma'}^{(2)}\ \tilde{\lambda}_{2l\Gamma \Gamma'}\ \hat{K}_{\Gamma \Gamma' 2}^{(1)}\Big]\nonumber \\
&+\sum_{\Gamma}d_{\Gamma}\ (d_{\Gamma}-1)\ {U}_{\Gamma \Gamma}^{(2)}\ \tilde{\lambda}_{2t\Gamma \Gamma}\ \hat{K}_{t\Gamma \Gamma}^{(2)}
+\sum_{(\Gamma,\Gamma')}d_{\Gamma}\ d_{\Gamma'}\ {U}_{\Gamma \Gamma'}^{(2)}\ \tilde{\lambda}_{2t\Gamma \Gamma'}\ \hat{K}_{t\Gamma \Gamma'}^{(2)}\,.
\label{equb24}
\end{align}
Here
\begin{align}
\hat{K}_{\Gamma \Gamma}^{(0)}&=\frac{1}{4}\ (d_{\Gamma}-1)\ U_{\Gamma\Gamma}^{(2)}\ \Big(\tilde{\lambda}_{2l\Gamma \Gamma}+2\ \tilde{\lambda}_{2t\Gamma \Gamma}\Big)\ M_{\Gamma\Gamma}\nonumber \\
&+\frac{1}{4}\sum_{\Gamma'(\neq \Gamma)}d_{\Gamma'}\ U_{\Gamma\Gamma'}^{(2)}\ \Big(\tilde{\lambda}_{2l\Gamma \Gamma'}+2\ \tilde{\lambda}_{2t\Gamma \Gamma'}\Big)\ M_{\Gamma\Gamma'}\,.
\label{equb25}
\end{align}
\begin{align}
\hat{K}_{\Gamma \Gamma'1}^{(1)}=-\frac{1}{16}\ U_{\Gamma\Gamma'}^{(2)}\ \Big(\tilde{\lambda}_{2l\Gamma \Gamma'}+2\ \tilde{\lambda}_{2t\Gamma \Gamma'}\Big)\ \Omega_{\Gamma \Gamma'}\,.
\label{equb26}
\end{align}
\begin{align}
\hat{K}_{\Gamma \Gamma'2}^{(1)}&=-U_{\Gamma\Gamma}^{(0)}\ M_{\Gamma\Gamma'}\ \tilde{\lambda}_{0\Gamma \Gamma}-U_{\Gamma'\Gamma'}^{(0)}\ M_{\Gamma'\Gamma}\ \tilde{\lambda}_{0\Gamma' \Gamma'}\nonumber \\
&+U_{\Gamma\Gamma'}^{(1)}\ \Omega_{\Gamma \Gamma'}\ \tilde{\lambda}_{1\Gamma \Gamma'}+\frac{1}{2}\ U_{\Gamma\Gamma'}^{(2)}\ W_{\Gamma \Gamma'}\ \tilde{\lambda}_{2t\Gamma \Gamma'}\nonumber \\
&+\frac{1}{2}\ \Big(U_{\Gamma\Gamma}^{(2)}\ \Xi_{\Gamma \Gamma'\Gamma}\ \tilde{\lambda}_{2l\Gamma \Gamma}+U_{\Gamma'\Gamma}^{(2)}\ \Xi_{\Gamma' \Gamma\Gamma}\ \tilde{\lambda}_{2l\Gamma' \Gamma}\Big)\nonumber \\
&+\frac{1}{2}\ \Big(U_{\Gamma\Gamma'}^{(2)}\ \Xi_{\Gamma \Gamma'\Gamma'}\ \tilde{\lambda}_{2l\Gamma \Gamma'}+U_{\Gamma'\Gamma'}^{(2)}\ \Xi_{\Gamma' \Gamma\Gamma'}\ \tilde{\lambda}_{2l\Gamma' \Gamma'}\Big)\nonumber \\
&-\frac{1}{2}\sum_{\Gamma''}d_{\Gamma''}\ \Big(U_{\Gamma\Gamma''}^{(2)}\ \Xi_{\Gamma \Gamma'\Gamma''}\ \tilde{\lambda}_{2l\Gamma \Gamma''}+U_{\Gamma'\Gamma''}^{(2)}\ \Xi_{\Gamma' \Gamma\Gamma''}\ \tilde{\lambda}_{2l\Gamma' \Gamma''}\Big)\,.
\label{equb27}
\end{align}
\begin{align}
\hat{K}_{t\Gamma \Gamma'}^{(2)}&=\frac{1}{4}\ \Big(U_{\Gamma\Gamma}^{(0)}\ M_{\Gamma\Gamma'}\ \tilde{\lambda}_{0\Gamma \Gamma}+U_{\Gamma'\Gamma'}^{(0)}\ M_{\Gamma'\Gamma}\ \tilde{\lambda}_{0\Gamma' \Gamma'}\Big)\nonumber \\
&-\frac{1}{4}\ \Big(4\ U_{\Gamma\Gamma'}^{(1)}\ \Omega_{\Gamma \Gamma'}\ \tilde{\lambda}_{1\Gamma \Gamma'}+U_{\Gamma\Gamma'}^{(2)}\ W_{\Gamma \Gamma'}\ \tilde{\lambda}_{2l\Gamma \Gamma'}-\frac{1}{4}\ U_{\Gamma\Gamma'}^{(2)}\ \Omega_{\Gamma \Gamma'}\ \tilde{\lambda}_{2t\Gamma \Gamma'}\Big)\nonumber \\
&-\frac{1}{8}\ \Big(U_{\Gamma\Gamma}^{(2)}\ \Xi_{\Gamma \Gamma'\Gamma}\ \tilde{\lambda}_{2t\Gamma \Gamma}+U_{\Gamma'\Gamma}^{(2)}\ \Xi_{\Gamma' \Gamma\Gamma}\ \tilde{\lambda}_{2t\Gamma' \Gamma}\Big)\nonumber \\
&-\frac{1}{8}\ \Big(U_{\Gamma\Gamma'}^{(2)}\ \Xi_{\Gamma \Gamma'\Gamma'}\ \tilde{\lambda}_{2t\Gamma \Gamma'}+U_{\Gamma'\Gamma'}^{(2)}\ \Xi_{\Gamma' \Gamma\Gamma'}\ \tilde{\lambda}_{2t\Gamma' \Gamma'}\Big)\nonumber \\
&+\frac{1}{8}\sum_{\Gamma''}d_{\Gamma''}\ \Big(U_{\Gamma\Gamma''}^{(2)}\ \Xi_{\Gamma \Gamma'\Gamma''}\ \tilde{\lambda}_{2t\Gamma \Gamma''}+U_{\Gamma'\Gamma''}^{(2)}\ \Xi_{\Gamma' \Gamma\Gamma''}\ \tilde{\lambda}_{2t\Gamma' \Gamma''}\Big)\,.
\label{equb28}
\end{align}
The expressions of $P_{\Gamma\Gamma'}$, $M_{\Gamma\Gamma'}$, $\Xi_{\Gamma \Gamma'\Gamma''}$, $\Omega_{\Gamma \Gamma'}$, and $W_{\Gamma\Gamma'}$ have been given in Appendix C.

\section{Momentum Distribution Function} 
In this Appendix we present the explicit expressions of $B_{LL'\sigma}({\epsilon}_{kn\sigma})$ and $C_{LL'\sigma}({\epsilon}_{kn\sigma})$ in Eq. (\ref{equ46b}) in the paramagnetic state, and derive the average quasiparticle weights (\ref{equ46ab}) and (\ref{equ48}). The numerator of the correlation correction to the momentum distribution $\langle{n}_{kn\sigma}\rangle$ is then presented in Eq. (\ref{equ46a}):
\begin{equation}
N\langle\tilde{O}_{i}^{\dagger}\tilde{n}_{kn\sigma}\tilde{O}_{i}\rangle_{0}=\sum_{\alpha \tau\  <LL'>} q_{\tau}^{(\alpha)}\ U_{LL'}^{(\alpha)2}\ \tilde{\lambda}_{\alpha\tau LL'}^{2}\ \Big(\hat{B}_{LL'n} (\bm{k})\ f(-\tilde{\epsilon}_{kn})-\hat{C}_{LL'n} (\bm{k})\ f(\tilde{\epsilon}_{kn})\Big)\,.
\label{eque1}
\end{equation}
The particle and hole contributions, $\hat{B}_{LL'n}(\bm{k})$ and $\hat{C}_{LL'n}(\bm{k})$ are expressed by Eq. (\ref{equ46b}):
\begin{equation}
\hat{B}_{LL'n} (\bm{k})=\vert u_{Ln}(\bm{k})\vert^{2}{B}_{L'L}({\epsilon}_{kn})+ \vert u_{L'n}(\bm{k})\vert^{2}{B}_{LL'}({\epsilon}_{kn})\,,
\label{eque2}
\end{equation}
\begin{equation}
\hat{C}_{LL'n} (\bm{k})=\vert u_{Ln}(\bm{k})\vert^{2}{C}_{L'L}({\epsilon}_{kn})+ \vert u_{L'n}(\bm{k})\vert^{2}{C}_{LL'}({\epsilon}_{kn})\,.
\label{eque3}
\end{equation}
Assuming that orbital $L$ belongs to an irreducible representation $\Gamma$, we obtain the expressions of the $B_{LL'}(\epsilon_{kn})$ and $C_{LL'}(\epsilon_{kn})$ as follows.
\begin{align}
B_{\Gamma \Gamma'}(\epsilon_{kn})=-\int_{0}^{\infty}{dt\ dt'e^{i(\epsilon_{c}-\epsilon_{kn})(t+t')}\ a_{\Gamma}(-t-t')\ b_{\Gamma}(t+t')\ a_{\Gamma'}(-t-t')}\,,
\label{eque4}
\end{align}
\begin{align}
C_{\Gamma \Gamma'}(\epsilon_{kn})=-\int_{0}^{\infty}{dt\ dt'e^{i(\epsilon_{c}+\epsilon_{kn})(t+t')}\ a_{\Gamma}(-t-t')\ b_{\Gamma}(t+t')\ b_{\Gamma'}(t+t')}\,.
\label{eque5}
\end{align}

The quasiparticle weight is obtained from the jump at the Fermi level: $Z_{{k_{F}}n}=\langle n_{kn\sigma}\rangle_{k_{{F}-}} -\langle n_{kn\sigma}\rangle_{k_{{F}+}}$. Here ${k_{{F}-}}\, ({k_{{F}+}})$ means the wavevector just below (above) the Fermi surface. According to Eq. (\ref{equ46}), it is given by
\begin{align}
Z_{{k_{F}}n}=1+\frac{\delta (N\langle\tilde{O}_{i}^{\dagger}\tilde{n}_{kn\sigma}\tilde{O}_{i}\rangle_{0})_{k_{F}}}{1+\langle{\tilde{O_i}^\dagger\tilde{O_i}}\rangle_0}\,.
\label{eque6}
\end{align}
Here the numerator of the correlation corrections is given by
\begin{align}
&\delta{(N\langle\tilde{O}_{i}^{\dagger}\tilde{n}_{kn\sigma}\tilde{O}_{i}\rangle_{0})}_{k_{F}}=-\sum_{\Gamma}\ \Bigl[d_{\Gamma}\ {U}_{\Gamma \Gamma}^{{(0)}^{2}}\ \tilde{\lambda}_{0\Gamma \Gamma}^{2}+2\ d_{\Gamma}\ (d_{\Gamma}-1)\ {U}_{\Gamma \Gamma}^{{(1)}^{2}}\ \tilde{\lambda}_{1\Gamma \Gamma}^{2} \nonumber \\
&\hspace{1cm}+\frac{1}{8}\ d_{\Gamma}\ (d_{\Gamma}-1)\ {U}_{\Gamma \Gamma}^{{(2)}^{2}}\ \Big(\tilde{\lambda}_{2l\Gamma \Gamma}^{2}+2\ \tilde{\lambda}_{2t\Gamma \Gamma}^{2}\Big)\Bigr]
\times \Big( \hat{B}_{\Gamma \Gamma n}(\bm{k}_{F})+ \hat{C}_{\Gamma \Gamma n}(\bm{k}_{F})\Big)\nonumber \\
&\hspace{1cm}-\sum_{(\Gamma,\Gamma')}d_{\Gamma}\ d_{\Gamma'}\Bigl[2\ {U}_{\Gamma \Gamma'}^{{(1)}^{2}}\ \tilde{\lambda}_{1\Gamma \Gamma'}^{2}+\frac{1}{8}\ \Big(\tilde{\lambda}_{2l\Gamma \Gamma'}^{2}+2\ \tilde{\lambda}_{2t\Gamma \Gamma'}^{2}\Big)\Bigr]
\times\Big( \hat{B}_{\Gamma \Gamma' n}(\bm{k}_{F})+ \hat{C}_{\Gamma \Gamma' n}(\bm{k}_{F})\Big)\,.
\label{eque7}
\end{align}
$\hat{B}_{\Gamma \Gamma' n} (\bm{k}_{F})$ and $\hat{C}_{\Gamma \Gamma' n} (\bm{k}_{F})$ are defined by Eqs. (\ref{eque2}) and (\ref{eque3}) in which $L$ and $L'$ have been replaced by their irreducible representations $\Gamma$ and $\Gamma'$. 

Taking average over the Fermi surface, we obtain the average quasiparticle weight $Z$, Eq. (\ref{equ46ab}) as follows.
\begin{align}
Z=1+\frac{\overline{\delta (N\langle\tilde{O}_{i}^{\dagger}\tilde{n}_{kn\sigma}\tilde{O}_{i}\rangle_{0})_{k_{F}}}}{1+\langle{\tilde{O_i}^\dagger\tilde{O_i}}\rangle_0}\,,
\label{eque10}
\end{align}
\begin{align}
&\overline{\delta{(N\langle\tilde{O}_{i}^{\dagger}\tilde{n}_{kn\sigma}\tilde{O}_{i}\rangle_{0})}}_{k_{F}}=-\sum_{\Gamma}\Bigl[\ d_{\Gamma}{U}_{\Gamma \Gamma}^{{(0)}^{2}}\ \tilde{\lambda}_{0\Gamma \Gamma}^{2}+2\ d_{\Gamma}\ (d_{\Gamma}-1)\ {U}_{\Gamma \Gamma}^{{(1)}^{2}}\ \tilde{\lambda}_{1\Gamma \Gamma}^{2} \nonumber \\
&\hspace{1cm}+\frac{1}{8}\ d_{\Gamma}\ (d_{\Gamma}-1)\ {U}_{\Gamma \Gamma}^{{(2)}^{2}}\ \Big(\tilde{\lambda}_{2l\Gamma \Gamma}^{2}+2\ \tilde{\lambda}_{2t\Gamma \Gamma}^{2}\Big)\Bigr]
 \times \Big( \bar{B}_{\Gamma \Gamma n}(\bm{k}_{F})+ \bar{C}_{\Gamma \Gamma n}(\bm{k}_{F})\Big)\nonumber \\
&\hspace{1cm}-\sum_{(\Gamma,\Gamma')}d_{\Gamma}\ d_{\Gamma'}\Bigl[2\ {U}_{\Gamma \Gamma'}^{{(1)}^{2}}\ \tilde{\lambda}_{1\Gamma \Gamma'}^{2}+\frac{1}{8}\ \Big(\tilde{\lambda}_{2l\Gamma \Gamma'}^{2}+2\ \tilde{\lambda}_{2t\Gamma \Gamma'}^{2}\Big) \Bigr]
\times\Big( \bar{B}_{\Gamma \Gamma' n}(\bm{k}_{F})+\bar{C}_{\Gamma \Gamma' n}(\bm{k}_{F})\Big)\,.
\label{eque11}
\end{align}
Here $\bar{B}_{\Gamma \Gamma'n} (\bm{k}_{F})$ and $\bar{C}_{\Gamma \Gamma'n} (\bm{k}_{F})$ are defined by 
\begin{equation}
\bar{B}_{\Gamma' \Gamma n} (\bm{k}_{F})=\overline{\vert u_{\Gamma n}(\bm{k}_{F})\vert^{2}}\ {B}_{\Gamma'\Gamma}({\epsilon}_{F})+ \overline{\vert u_{\Gamma' n}(\bm{k}_{F})\vert^{2}}\ {B}_{\Gamma\Gamma'}({\epsilon}_{F})\,,
\label{eque12}
\end{equation}
\begin{equation}
\bar{C}_{\Gamma \Gamma'n} (\bm{k}_{F})=\overline{\vert u_{\Gamma n}(\bm{k}_{F})\vert^{2}}\ {C}_{\Gamma' \Gamma}({\epsilon}_{F})+ \overline{\vert u_{\Gamma' n}(\bm{k}_{F})\vert^{2}}\ {C}_{\Gamma \Gamma'}({\epsilon}_{F})\,.
\label{eque13}
\end{equation}
The average amplitude of eigenvector $\overline{\vert u_{\Gamma n}(\bm{k}_{F})\vert^{2}}$ is obtained as follows.
\begin{align}
\overline{\vert u_{\Gamma n}(\bm{k}_{F})\vert^{2}}&=\frac{\sum_{kn}^{\epsilon_{F}\langle\epsilon_{kn}\langle\epsilon_{F}+\Delta}\vert u_{\Gamma n}(\bm{k})\vert^{2}}{\sum_{kn}^{\epsilon_{F}\langle\epsilon_{kn}\langle\epsilon_{F}+\Delta}} 
=\frac{\rho_{\Gamma}(\epsilon_{F})}{\rho (\epsilon_{F})}\,.
\label{eque15}
\end{align}
Here $\rho_{\Gamma}(\epsilon_{F})\,(\ \rho(\epsilon_{F}))$ is the partial (total) density of states at the Fermi level $\epsilon_{F}$. 

The projected momentum distribution function (MDF) is defined by 
\begin{equation}
\langle n_{kL\sigma}\rangle=\sum_{n}\langle n_{kn\sigma}\rangle\vert u_{Ln\sigma}(\bm{k})\vert^{2}\,.
\label{eque14}
\end{equation}
Using the formula (\ref{equ14}), we obtain the expression of the projected MDF, Eq.(\ref{equ47}): 
\begin{equation}
\langle n_{kL\sigma}\rangle=f(\tilde{\epsilon}_{kL\sigma})+\frac{N\langle\tilde{O}_{i}^{\dagger}\tilde{n}_{kL\sigma}\tilde{O}_{i}\rangle_{0}}{1+\langle{\tilde{O_i}^\dagger\tilde{O_i}}\rangle_0}\,.
\label{eque47}
\end{equation}
The correlation correction of the projected MDF at the rhs is expressed as follows after taking the average over $\bm{k}$ with constant energy $\epsilon_{kL}$ as in Eq. (\ref{eque15}).
\begin{align}
N\langle\tilde{O}_{i}^{\dagger}\tilde{n}_{kL\sigma}\tilde{O}_{i}\rangle_{0}&=\frac{D\ \rho_{\Gamma}(\epsilon_{kL})} {\rho(\epsilon_{kL})}\sum_{\Gamma'}\ \Bigl[d_{\Gamma'}\ {U}_{\Gamma' \Gamma'}^{{(0)}^{2}}\ \tilde{\lambda}_{0\Gamma' \Gamma'}^{2}+2\ d_{\Gamma'}\ (d_{\Gamma'}-1)\ {U}_{\Gamma' \Gamma'}^{{(1)}^{2}}\ \tilde{\lambda}_{1\Gamma' \Gamma'}^{2} \nonumber \\
&+\frac{1}{8}\ d_{\Gamma'}\ (d_{\Gamma'}-1)\ {U}_{\Gamma' \Gamma'}^{{(2)}^{2}}\Big(\tilde{\lambda}_{2l\Gamma' \Gamma'}^{2}+2\ \tilde{\lambda}_{2t\Gamma' \Gamma'}^{2}\Big) \Bigr]\nonumber \\
&\times\frac{\rho_{\Gamma'}(\epsilon_{kL})}{\rho(\epsilon_{kL})}\ \Big( B_{\Gamma' \Gamma'}(\epsilon_{kL})\ f(-\tilde{\epsilon}_{kL})- C_{\Gamma' \Gamma'}(\epsilon_{kL})\ f(\tilde{\epsilon}_{kL})\Big)\nonumber \\
&+\frac{D\ \rho_{\Gamma}(\epsilon_{kL})} {\rho(\epsilon_{kL})}\sum_{(\Gamma',\Gamma'')}d_{\Gamma'}\ d_{\Gamma''}\Bigl[2\ {U}_{\Gamma' \Gamma''}^{{(1)}^{2}}\ \tilde{\lambda}_{1\Gamma' \Gamma''}^{2}+\frac{1}{8}\ \Big(\tilde{\lambda}_{2l\Gamma' \Gamma''}^{2}+2\ \tilde{\lambda}_{2t\Gamma' \Gamma''}^{2}\Big) \Bigr]\nonumber \\
& \times\Bigl[\frac{\rho_{\Gamma'}\ (\epsilon_{kL})}{\rho(\epsilon_{kL})}\ \Big( B_{\Gamma'' \Gamma'}(\epsilon_{kL})\ f(-\tilde{\epsilon}_{kL})- C_{\Gamma'' \Gamma'}(\epsilon_{kL})\ f(\tilde{\epsilon}_{kL})\Big)\nonumber \\
&+\frac{\rho_{\Gamma''}(\epsilon_{kL})}{\rho(\epsilon_{kL})}\ \Big( B_{\Gamma' \Gamma''}(\epsilon_{kL})\ f(-\tilde{\epsilon}_{kL})- C_{\Gamma' \Gamma''}(\epsilon_{kL})\ f(\tilde{\epsilon}_{kL})\Big)\Bigr]\,.
\label{eque17}
\end{align}
This is the explicit expression of the numerator of the second term of Eq. (\ref{equ47}). 

With use of Eqs. (\ref{eque47}) and (\ref{eque17}), the partial quasiparticle weight $Z_{L}$ is given by Eq. (\ref{equ48}):
\begin{align}
Z_{L}=1+\frac{\overline{\delta(N\langle\tilde{O}_{i}^{\dagger}\tilde{n}_{kL\sigma}\tilde{O}_{i}\rangle_{0})}_{k_{F}}}{1+\langle{\tilde{O_i}^\dagger\tilde{O_i}}\rangle_0}\,,
\label{eque48}
\end{align}
and the explicit expression of the numerator of the correlation correction is given as follows.
\begin{align}
&\overline{\delta{(N\langle\tilde{O}_{i}^{\dagger}\tilde{n}_{kL\sigma}\tilde{O}_{i}\rangle_{0})}}_{k_{F}}=-\sum_{\Gamma}\ \Bigl[d_{\Gamma}\ {U}_{\Gamma \Gamma}^{{(0)}^{2}}\ \tilde{\lambda}_{0\Gamma \Gamma}^{2}+2\ d_{\Gamma}\ (d_{\Gamma}-1)\ {U}_{\Gamma \Gamma}^{{(1)}^{2}}\ \tilde{\lambda}_{1\Gamma \Gamma}^{2} \nonumber \\
&\hspace{1cm}+\frac{1}{8}\ d_{\Gamma}\ (d_{\Gamma}-1)\ {U}_{\Gamma \Gamma}^{{(2)}^{2}}\ \Big(\tilde{\lambda}_{2l\Gamma \Gamma}^{2}+2\ \tilde{\lambda}_{2t\Gamma \Gamma}^{2}\Big)\Bigr]
 \times\frac{\rho_{\Gamma}(\epsilon_{F})}{\rho(\epsilon_{F})}\ \Big( B_{\Gamma \Gamma}(\epsilon_{F})+ C_{\Gamma \Gamma}(\epsilon_{F})\Big)\nonumber \\
&\hspace{1cm}-\sum_{(\Gamma,\Gamma')}d_{\Gamma}\ d_{\Gamma'}\Bigl[2\ {U}_{\Gamma \Gamma'}^{{(1)}^{2}}\ \tilde{\lambda}_{1\Gamma \Gamma'}^{2}+\frac{1}{8}\ \Big(\tilde{\lambda}_{2l\Gamma \Gamma'}^{2}+2\ \tilde{\lambda}_{2t\Gamma \Gamma'}^{2}\Big)\Bigr]\nonumber \\
&\hspace{1cm} \times\Bigl[\frac{\rho_{\Gamma}(\epsilon_{F})}{\rho(\epsilon_{F})}\ \Big( B_{\Gamma' \Gamma}(\epsilon_{F})+C_{\Gamma' \Gamma}(\epsilon_{F})\Big)
+\frac{\rho_{\Gamma'}(\epsilon_{F})}{\rho(\epsilon_{F})}\ \Big( B_{\Gamma \Gamma'}(\epsilon_{F})+ C_{\Gamma \Gamma'}(\epsilon_{F})\Big)\Bigr]\,.
\label{eque18}
\end{align}
%


\end{document}